# Imaging the Photochemical Ring-Opening of 1,3-Cyclohexadiene by Ultrafast Electron Diffraction

T. J. A. Wolf[1,*], D. M. Sanchez[1,2], J. Yang[1,3], R. M. Parrish[1,2], J. P. F. Nunes[4,5], M. Centurion[5], R. Coffee[3], J. P. Cryan[1], M. Gühr[1,6], K. Hegazy[1,7], A. Kirrander[8], R. K. Li[3], J. Ruddock[9], X. Shen[3], T. Veccione[3], S. P. Weathersby[3], P. M. Weber[9], K. Wilkin[5], H. Yong[9], Q. Zheng[3], X. J. Wang[3,*], M. P. Minitti[3,*], T. J. Martínez[1,2,*]

[1]Stanford PULSE Institute, SLAC National Accelerator Laboratory, Menlo Park, USA.

[2]Department of Chemistry, Stanford University, Stanford, USA.

[3]SLAC National Accelerator Laboratory, Menlo Park, USA.

[4]Department of Chemistry, University of York, Heslington, York, UK.

[5]Department of Physics and Astronomy, University of Nebraska-Lincoln, Lincoln, USA.

[6]Institut für Physik und Astronomie, Universität Potsdam, Potsdam, Germany.

[7]Department of Physics, Stanford University, Stanford, USA.

[8]EaStCHEM, School of Chemistry, University of Edinburgh, Edinburgh EH9 3FJ, United Kingdom.

[9]Department of Chemistry, Brown University, Providence, USA.

**Abstract:**

The ultrafast photoinduced ring-opening of 1,3-cyclohexadiene constitutes a textbook example of electrocyclic reactions in organic chemistry and a model for photobiological reactions in vitamin D synthesis. Here, we present direct and unambiguous observation of the ring-opening reaction path on the femtosecond timescale and sub-Ångström length scale by megaelectronvolt ultrafast electron diffraction. We follow the carbon-carbon bond dissociation and the structural opening of the 1,3-cyclohexadiene ring by direct measurement of time-dependent changes in the distribution of interatomic distances. We observe a substantial acceleration of the ring-opening motion after internal conversion to the ground state due to steepening of the electronic potential gradient towards the product minima. The ring-opening motion transforms into rotation of the terminal ethylene groups in the photoproduct 1,3,5-hexatriene on the sub-picosecond timescale. Our work demonstrates the potential of megaelectronvolt ultrafast electron diffraction to elucidate photochemical reaction paths in organic chemistry.

**Main text**

The photoinduced ring-opening of 1,3-cyclohexadiene (CHD) yielding 1,3,5-hexatriene (HT) is a prototypical electrocyclic reaction that provides a model system for understanding vitamin D generation.[1,2] As with any photoinduced electrocyclic reaction, the ring-opening of CHD is characterized by concerted rearrangement of single and double bonds and strong stereoselectivity. The latter is well-described by the celebrated Woodward-Hoffmann rules.[3] Analogous reactions enable many otherwise difficult transformations in organic synthesis[4] and serve as the basis for many molecular switches.[5]

After photoexcitation at 267 nm to the first excited singlet state ($S_1$, see Fig. 1), the ring-opening reaction proceeds by nonradiative relaxation through a conical intersection (CI) to the ground state ($S_0$) of the reaction product HT.[1,6–8] Near the CI, correlated motion of electrons and nuclei leads to efficient nonadiabatic transitions from $S_1$ to $S_0$. The energy of the absorbed photon initially alters only the electronic wavefunction, but is rapidly translated into a rearrangement of atoms, *i.e.* a photochemical reaction.

In the case of CHD ring-opening, a ring of four single and two double C-C bonds is transformed into an alternating chain of three double and two single C-C bonds. There are three structural isomers of the ring-opened HT photoproduct (see Fig. 1 and Supplementary Discussion 1), differing by torsions about the C-C single bonds. The barriers separating these isomers are low (≈0.2eV) compared to the excess energy from the absorbed photon that is available to the nuclei after relaxation to $S_0$ (≈3.8eV).[6] Therefore, the ground state nuclear wavepacket can be expected to evolve into a mixture of all three isomers. Although previous investigations with ps time resolution have shown that ground state equilibration takes place within several to hundreds of ps,[9–11] direct observation of the atomic displacements in both the initial ring-opening and the earliest sub-picosecond ground state isomerization dynamics has yet to be achieved.[1,6]

The ring-opening has been studied extensively in the gas phase by optical and x-ray spectroscopic methods (see Refs.[1,6,12–15] and Refs. cited therein). Due to their preferential sensitivity to changes in the electronic wavefunction, these experiments reveal timescales for population transfer between electronic states through CIs,[16,17] but cannot directly observe structural dynamics on atomic space and time scales. Time-resolved vibrational spectroscopies,[18] which in principle exhibit such sensitivity, are intrinsically insensitive to dynamics along steeply repulsive potentials like the CHD ring-opening path. Pioneering time-resolved x-ray and electron diffraction studies have made impressive progress towards resolving ultrafast structural dynamics of isolated organic molecules,[9,19–23] but until now, have fallen short of either the sub-Å spatial or femtosecond temporal resolution needed to follow photochemical reaction dynamics in these systems. Early influential non-relativistic electron diffraction studies in molecular crystals have provided unambiguous evidence of ultrafast structural dynamics.[24–26] However, the crystalline environment can have a significant influence on the observed dynamics due to constraints on large amplitude motions from crystal packing effects.[27,28]

Recently, seminal megaelectronvolt (MeV) ultrafast electron diffraction (UED) studies of increasing complexity, from rotational dynamics in diatomic $N_2$[29] to vibrational dynamics in diatomic $I_2$[30] and dissociative dynamics in penta-atomic $CF_3I$,[31] demonstrated the resolution in space and time required to elucidate ultrafast structural dynamics outside crystalline environments. In most of these cases, heavy atoms were involved in order to achieve the necessary signal to noise ratio.

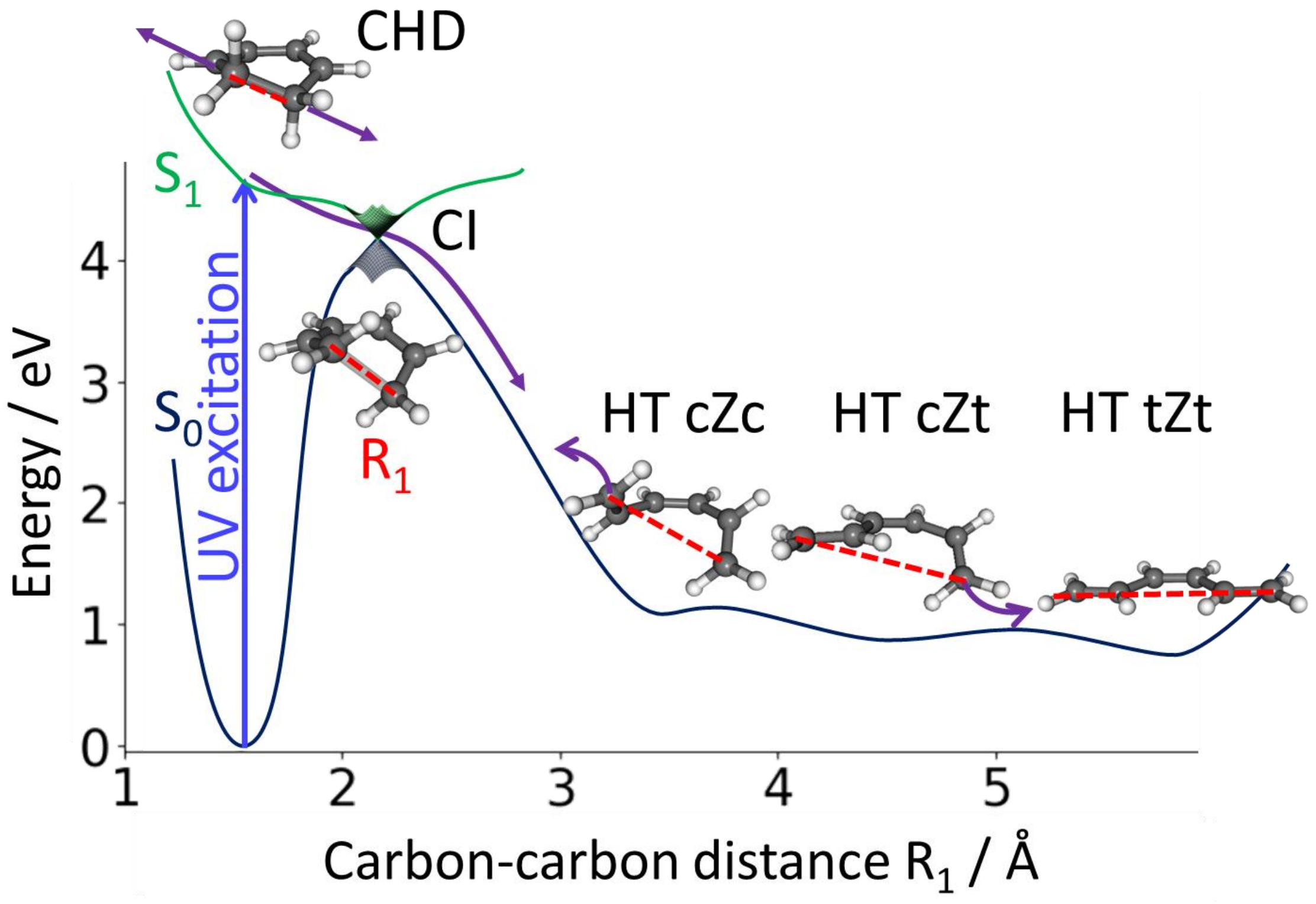

**Figure 1: Schematic of the photoinduced ring-opening reaction of 1,3 cyclohexadiene (CHD).** CHD is photoexcited from its closed-ring ground state ($S_0$) energy minimum to an excited state ($S_1$). It evolves along the ring-opening coordinate (indicated by purple arrows) through a conical intersection (CI) by elongation of the C-C distance $R_1$ (red) to the $S_0$ potential energy surface region of 1,3,5-hexatriene (HT). The molecule transforms from a ring containing two conjugated double bonds to a chain of three conjugated ethylene subunits. Twisting about the newly formed single C-C bonds connects three isomers of HT (cZc, cZt, and tZt) via low barriers. The depicted potential energy curves are based on calculated energies at minimum, barrier, and CI geometries (see Supplementary Discussion 2).

In the following, we show that MeV UED allows us to directly observe both the excited state reaction path and subsequent ground state isomerization dynamics for ring-opening in CHD, with sub-Å/femtosecond resolution in space/time for transient changes of atomic distances. Our work resolves atomic motion on femtosecond timescales for the photochemistry of a polyatomic organic molecule containing exclusively light elements with small scattering cross-sections. We believe this is a milestone in enabling MeV UED for general investigations of ultrafast gas phase organic photochemistry. We cover a momentum transfer space (see Fig. 2b and 2d), which is similar to previous ps time-resolved electron diffraction studies[9,19,20,22] but considerably larger than previous fs time-resolved x-ray scattering studies.[21] We approach the maximum momentum transfer range which was recently identified as reasonable for the investigation of structural dynamics in CHD.[32] Therefore, as opposed to the previous x-ray scattering studies, our diffraction data permit reliable transformation into real-space atomic pair distribution functions (PDFs) without any input from theory or simulation. This allows us to directly compare our data to *ab-initio* simulations of the reaction dynamics. As the theory and experiment are completely independent of each other, the successful comparison provides a compelling test of both.

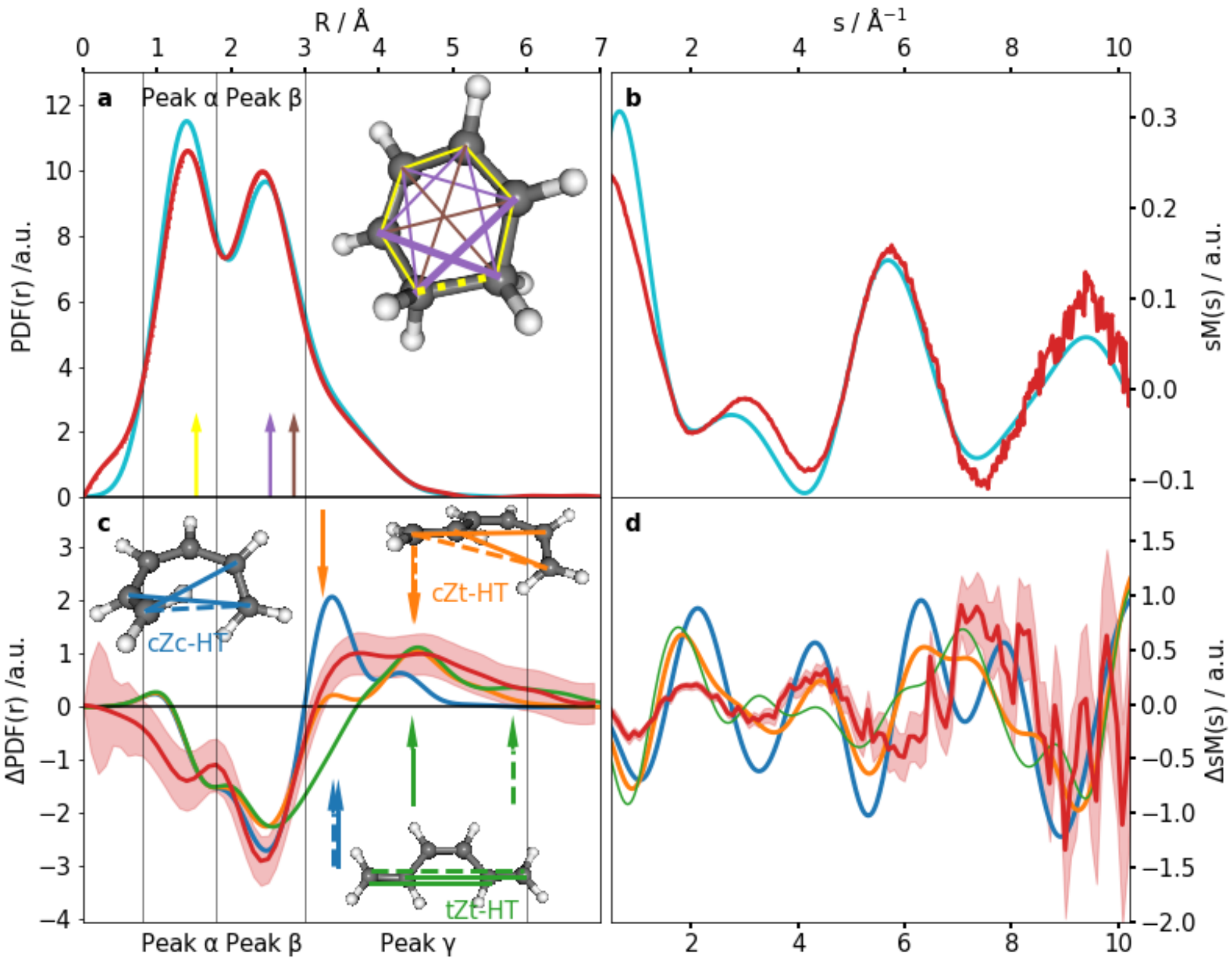

**Figure 2: Comparison of experimental and simulated atomic pair distribution functions (PDF).** a) Experimental (red) and simulated (light blue) steady-state PDFs of 1,3-cyclohexadiene (CHD) based on an optimized *ab-initio* geometry (see inset and Supplementary Discussion 2-3). The two peaks (α, β, see text) of the PDFs correspond to C-C bond distances (yellow) and C-C distances across the ring (purple, brown). The distances $R_1$ and $R_2$, which change significantly during ring-opening, are highlighted by dotted and solid bold lines in the inset. b) Simulated (light blue) and experimental (red) modified molecular diffraction signals, sM(s), for the CHD reactant in momentum-transfer space. c) Steady state simulated difference PDFs (ΔPDFs, $PDF_{HT}$-$PDF_{CHD}$) for the three 1,3,5-hexatriene (HT) isomers, cZc-HT (blue), cZt-HT (orange), and tZt-HT (green). For comparison, an experimental ΔPDF at t=0.55 ps is shown (red). The distances $R_1$ and $R_2$ (dotted and solid bold lines, respectively) are marked in the geometries and shown as arrows. d) Structural information from c) in momentum-transfer space as difference sM(s) (ΔsM(s)). Error bars represent a 68 % confidence interval obtained from bootstrap analysis.[33]

## Results

Figure 2 shows steady-state structural information of CHD in real space (PDF in Fig. 2a) and momentum transfer space (modified molecular diffraction, sM(s), Fig. 2b), respectively. The experimental results are compared with a simple simulation based on an *ab initio*-computed ground state minimum geometry of CHD (see Supplementary Discussion 2 and 3). Experimental and simulated steady-state diffraction signals are in reasonable agreement. The minor observed differences can be ascribed to the method used to subtract the atomic background in the experimental data and the approximation of the ground state nuclear wavefunction by a single geometry (see Supplementary Discussion 3). The corresponding real-space PDFs in Fig. 2a (see Supplementary Discussion 4) exhibit two peaks at 1.4 Å (peak α) and 2.4 Å (peak β). Peak α refers to nearest-neighbor C-C bond distances and peak β is associated with two types of C-C distances across the CHD ring (see inset of Fig. 2a). An additional shoulder at peak β towards larger distances is due to C-H pair correlations. Because the intensity scales with the product of the nuclear charges, the C-H contributions are substantially weaker and H-H contributions are negligible. Accordingly, we will focus on C-C distances below.

The red lines in Figures 2c and 2d respectively show real-space (ΔPDF, see Supplementary Discussion 5) and momentum transfer space (ΔsM) experimental difference signals (time-dependent signature minus static signature) at 0.55 ps time delay, when the ring-opening is complete. They are compared with simulated steady state difference signatures for each of the three HT isomers. We focus our analysis on the real-space representation. In ΔPDFs, the change of an individual C-C distance appears as a pair of correlated features: a negative contribution at the initial value in the CHD reactant and a positive contribution corresponding to its new value at the specific time delay. The ΔPDF is governed by contributions from the C-C distances marked in Fig. 2a and 2c, which undergo substantial changes, whereas contributions to ΔPDF from other distances are weak. The negative signatures coincide with peaks α and β in the static PDF of CHD and refer to the broken C-C bond ($R_1$, marked by bold dotted lines in the inset of Fig. 2a) and C-C distances across the ring ($R_2$, marked by bold solid lines in Fig. 2a), respectively. Moreover, the positive feature (peak γ) between 3 Å and 6 Å, refers to the corresponding C-C distances in HT (see insets of Fig. 2c). Since these distances are larger than those in CHD, the positive peak is direct and unambiguous proof of photoinduced ring-opening.

We compare the experimental ΔPDF in Fig. 2c to simple simulations based on the three isomeric minimum geometries of HT. There is qualitative agreement, with all three ground state isomers and the experimental ΔPDF exhibiting negative signatures below 3 Å and positive signatures beyond 3 Å. However, the experimental ΔPDF clearly does not correspond to a single HT isomer. Furthermore, there is complex structure in the positive signal beyond 3 Å that cannot obviously be attributed to a combination of the equilibrium isomer structures. This is most likely caused by substantial broadening of the nuclear wavepacket due to the large amount of kinetic energy redistributed into nuclear degrees of freedom as the molecule returns to the ground state. We, therefore, refrain from attempts to retrieve transient structures from the experimental data[9,20,24–26] and, instead, interpret them in comparison to explicit wavepacket simulations (see below).

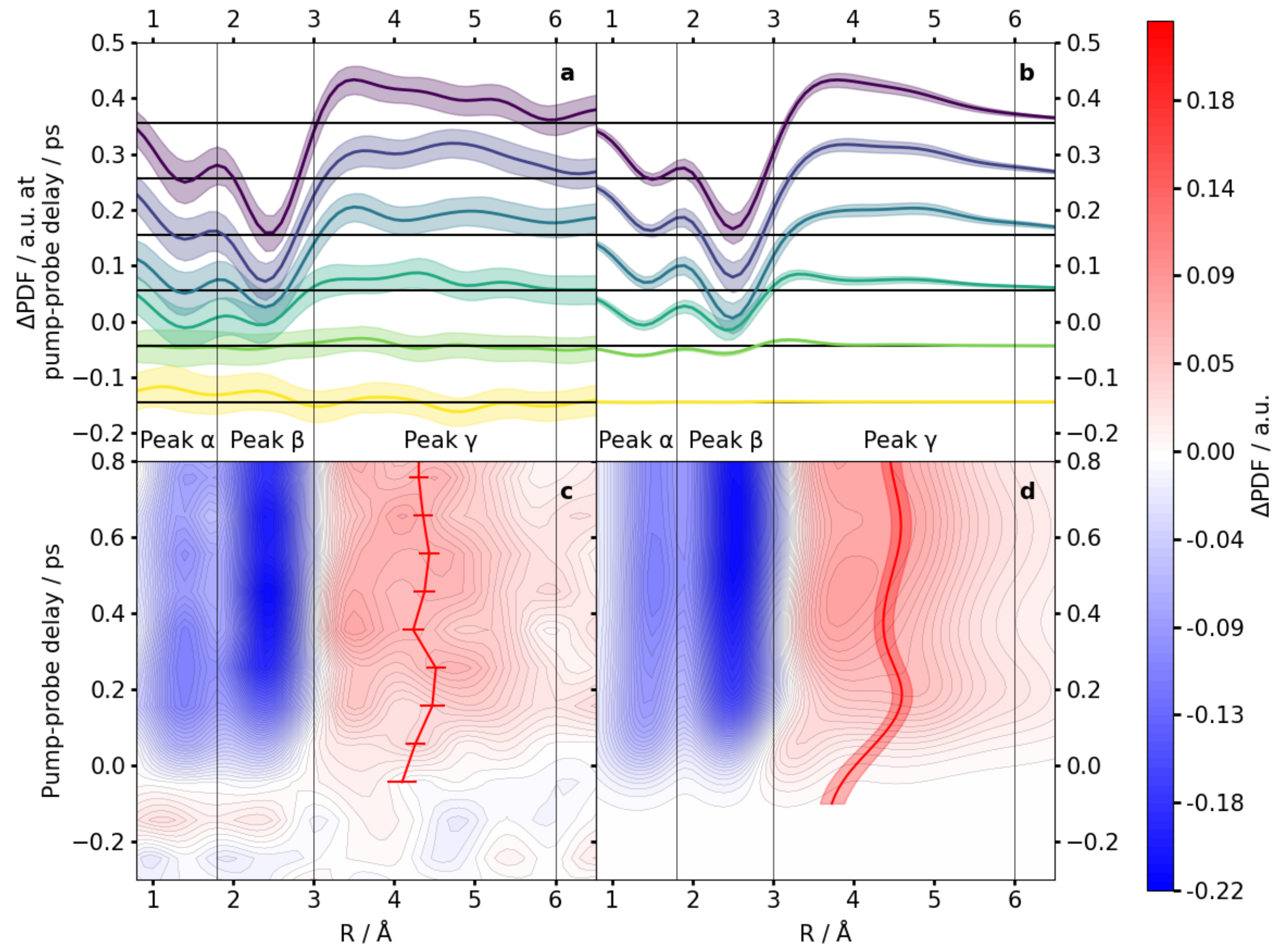

**Figure 3: Comparison of experimental (a,c) and simulated (b,d) time-dependent difference pair distribution functions (ΔPDF).** a,b) ΔPDFs at different time delays after photoexcitation and c,d) false-color plots of ΔPDF over the whole investigated time window. The plots are separated into peak regions α-γ. The center-of-mass position in peak area γ is shown as red curves in c, d. Simulated ΔPDFs are based on *ab-initio* multiple spawning simulations (see methods) and are convolved with a temporal Gaussian to account for the experimental response function. Error bars represent a 68 % confidence interval obtained from bootstrap analysis.[33] For the simulations, these error bars reflect convergence with respect to initial condition sampling.

Figures 3a and 3c show experimental ΔPDFs at different delay times after photoexcitation (t=0) and a false-color surface plot of the whole dataset. Convolved with the finite instrument response function (160 fs, see Supplementary Discussion 6), they exhibit time-dependent relative intensity changes due to the underlying structural evolution of the photoexcited molecules. The relative amplitudes of peaks α and β are similar for early delays, but peak β achieves almost twice the amplitude of peak α at later delays (see Fig. 3a). Fitting the changes in relative amplitudes with error functions reveals a delayed rise time (with respect to peak α) for peaks β (70 ± 30 fs) and γ (80 ± 40 fs).

The delayed onset of peaks β and γ directly reflects the ring-opening structural dynamics. During the ring-opening, positive contributions increase in both regions due to lengthening of $R_1$ and $R_2$. As shown in Supplementary Figure 1, the delayed rise of peak β can be attributed to the positive contribution from lengthening of $R_1$ (with amplitude moving from peak $\alpha$ to peak β) compensating for the negative contribution in this region from lengthening of $R_2$ (with amplitude moving from peak β to peak γ). The delayed rise of peak γ marks the time when $R_1$ reaches values corresponding to the HT isomers (see Fig. 1) and the ring-opening is complete.

To further elucidate the observed structural changes in the ΔPDFs, we compare our experimental data to *ab-initio* multiple spawning (AIMS) simulations[34] at the α-CASSCF(6,4)/6-31G*[35] level of theory (see methods). The computed ΔPDFs show excellent agreement with experiment (see Fig. 3 and Supplementary Fig. 2). The AIMS simulations also exhibit a delay in the onset of peaks β and γ (Fig. 4) relative to onset of peak α.

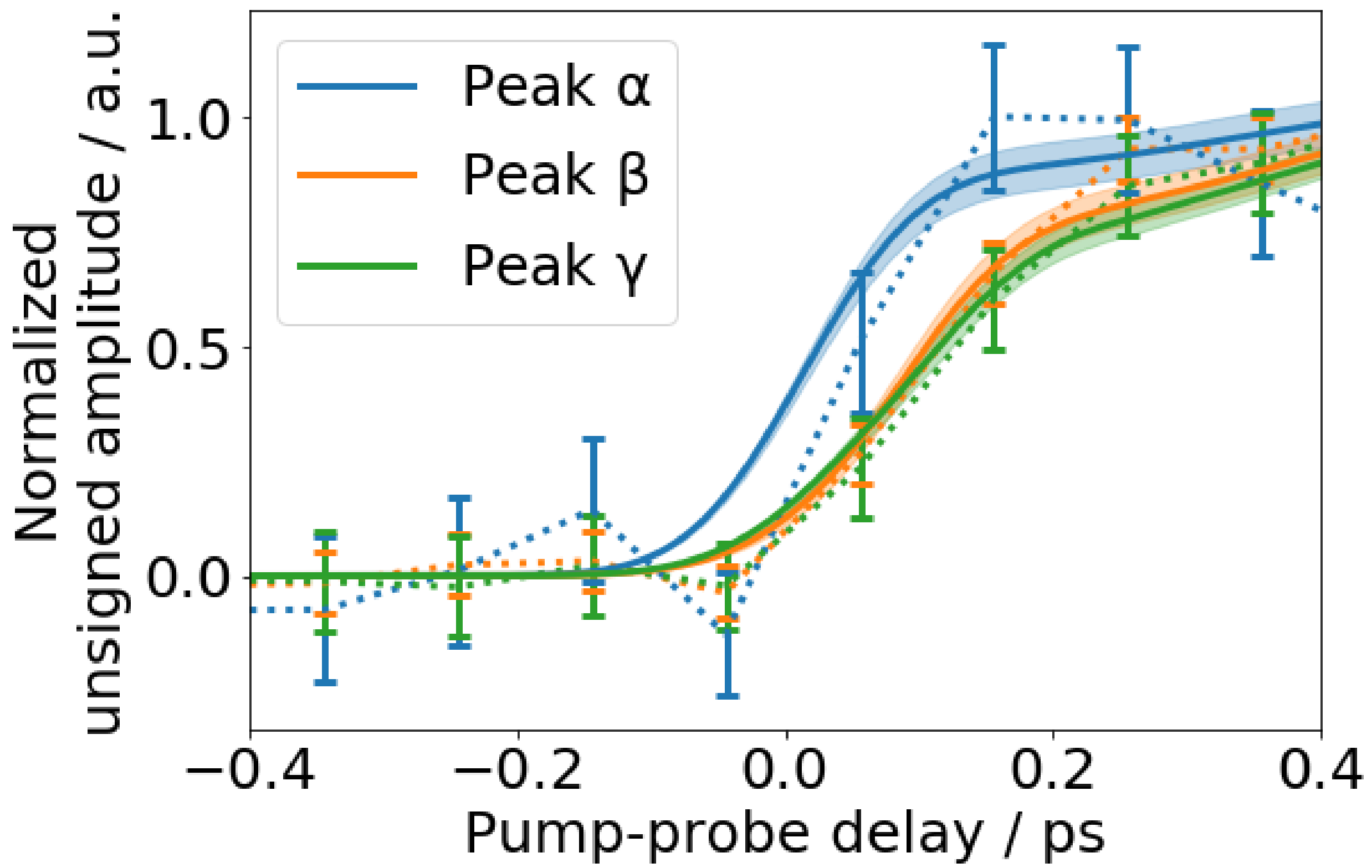

**Figure 4: Comparison between experimental and simulated delay in rise time between peaks α, β, and γ.** The areas of peak α (0.8 Å – 1.8 Å, blue), peak β (1.8 Å – 3.0 Å, orange), and peak γ (3.0 Å – 6.0 Å, green) in experimental (dotted lines) and simulated (solid lines) ΔPDFs as shown in Fig. 3 are integrated and normalized to the maximum unsigned amplitude. Error bars represent a 68 % confidence interval obtained from bootstrap analysis.[33]

We see breathing-like behavior of peak γ (modulating its width and peak value) in both experimental and simulated ΔPDFs. We quantify this by denoting the time-dependent center-of-mass of peak γ (red curves in Figs. 3c and d), which shows a maximum displacement at 0.25 ps (simulation: 0.2 ps), followed by a minimum at 0.35 ps (0.36 ps), and a weaker second maximum at 0.54 ps (0.59 ps). Comparison with the signatures of the HT isomers in Fig. 2b suggests a coherent oscillation of population between cZc-like and cZt/tZt-like geometries (see Supplementary Discussion 13). Separation of contributions from specific C-C distances in the AIMS dynamics and analysis of the time-dependent isomer distribution (see

Supplementary Figs. 3 and 14-16) indeed proves that the oscillation originates from $R_1$ and $R_2$ due to an evolving isomeric composition.

**Discussion**

As mentioned above, peaks α, β, and γ result from the interplay of positive and negative contributions due to structural changes after photoexcitation. The earliest, negative contribution, peak α, results from the increase in $R_1$ due to ring-opening. Unlike observables from time-resolved electronic spectroscopy, which usually exhibit features from the quasi-instantaneous response of the molecule's *electrons* to the photoexcitation, the onset of peak α does not mark the time of photoexcitation. Instead, it represents the onset of substantial excited state *nuclear* motion as a response to the photoexcitation, i.e. departure of the molecule from the Franck-Condon region. Thus, some delay is expected between photoexcitation and the onset of peak α. Based on our simulations, this delay is estimated to be 40 fs. Due to relatively small displacements, we are not able to observe other known Franck-Condon active degrees of freedom like the C-C bond alternation (which could be expected to respond more quickly to electronic excitation).

Depletion in the peak β region arises from lengthening of the $R_2$ distances in the CHD reactant. As discussed above and shown in Supplementary Fig. 1, the delay in the onset of this depletion is due to the simultaneous lengthening of $R_1$ distances (negative signal in peak α and positive signal in peak β), which partially compensates for the lengthening $R_2$ distances (negative signal in peak β and positive signal in peak γ). Coincidentally, the $R_2$ distance in CHD (2.4 Å) is very close to the $R_1$ distance of the open-minimum energy conical intersection geometry (2.2 Å, $S_1/S_0$ MECI (Open) in Supplementary Fig. 9) and the $R_1$ distances where we observe the majority of electronic transitions in the AIMS simulations. Therefore, the rise time of peak β marks the departure of the nuclear wavepacket from the excited state. After accounting for the delay between photoexcitation and the onset of peak α (40 fs), the delay between peak α and peak β (70 ± 30 fs) yields an excited state depopulation timescale of 110 ± 30 fs.

This is in agreement with both excited state lifetimes previously obtained from time-resolved electronic spectroscopy experiments (136 fs, 130 fs, 142 fs, and 139 fs),[13,14,36,37] and AIMS simulations (139 ± 25 fs, see Supplementary Figure 12).

Peak γ corresponds to photoexcited population with $R_1$ and $R_2$ distances beyond 2.4 Å. Its rise, thus, exclusively shows structural dynamics on the electronic ground state. This is confirmed by separating excited state and ground state contributions to the simulated ΔPDFs in Supplementary Figs. 4 and 5. Thus, the ring-opening is initiated in the excited state and completed in the ground state, whereas any following isomerization dynamics take place on the ground state. The onset time of peak γ is approximately constant over its whole range between 3 Å and 6 Å. Thus, as opposed to the $R_1$ increase from 1.4 Å to beyond 2.4 Å resulting in the delay between peaks α and γ, our time resolution is insufficient to fully time-resolve increases in $R_1$ from 3 Å to 6 Å (see Supplementary Figs. 1, 3, 4, and 5). Hence, the nuclear wavepacket must experience a substantial acceleration upon internal conversion to the ground state. This can be anticipated from the quantum chemical potential sketched in Fig. 1. The conical intersection is only slightly lower in potential energy than the Franck-Condon region. Therefore, the majority of the absorbed photoenergy is released *after* internal conversion to the ground state into the $R_1$ degree of freedom. In the reaction product HT, this kinetic energy is rapidly converted into twisting motions around the two newly formed single bonds (see Fig. 1) connecting the different HT isomeric minima. The kinetic energy released into the torsional degrees of freedom is so large that the barriers between the HT isomer minima do not play a dominant role. This results in quasi-free rotation of the terminal ethylene groups around the conjugated bonds (see Supplementary Movie 2). The rotation translates to coherent oscillation of the center of mass in peak γ, since the distances probed by this region are a direct reflection of the HT isomers (see Fig. 2 and Supplementary Discussion 13).

Signatures of the rotation of terminal ethylene groups in the HT photoproduct are observed here for the first time. Vacuum ultraviolet and soft x-ray electronic spectroscopic investigations showing signatures of the HT photoproduct do not seem to be sensitive to its isomerization dynamics.[12,13] The product distribution resulting from HT isomerization dynamics has been observed by previous ps UED experiments, but the isomerization mechanism could not be resolved.[9] Other UED studies observed formation of the tZt-HT isomer to take place within 20 ps.[11] In contrast, our study shows both theoretical and experimental evidence that the tZt-HT isomer is already accessed within 0.25 ps after photoexcitation. Signatures of the three HT isomers can be distinguished in solution phase transient absorption spectra.[38] However, in solution environments the solvent is expected to quickly dissipate vibrational excess energy from the solute. The resulting vibrational cooling renders the HT isomerization barrier heights significant enough to prevent terminal ethylene rotation, substantially altering the isomerization dynamics.

In conclusion, by following the femtosecond changes in the atomic distances $R_1$ and $R_2$, we demonstrate the first direct observation of the photoinduced structural ring-opening in isolated CHD, a model for the photosynthesis of previtamin $D_3$. Moreover, we find that the majority of the excess kinetic energy is released after return to the electronic ground state into a specific motion of the HT photoproduct - the quasi-free rotation of the terminal ethylene groups around the conjugated C-C bonds. This results in coherent oscillations in the atomic pair distribution signatures of the HT photoproduct, reflecting a highly non-equilibrium time-dependent oscillatory distribution of HT isomers. Our results showcase the promise of megaelectronvolt ultrafast electron diffraction for the study of ultrafast photochemistry and photobiology.

**Methods:**

**Megaelectronvolt gas phase ultrafast electron diffraction:** The gas-phase ultrafast electron diffraction (UED) apparatus is described in detail elsewhere[29,39]. In short, we use the 800 nm output of a Ti:Sapphire laser system operated at 180 Hz repetition rate and separate two beam paths. Pulses in both beam paths are frequency tripled. The pulses of the probe beam path are directed onto the photocathode of an RF gun and eject an ultrashort pulse containing ~$10^4$ electrons. The electrons are rapidly accelerated in a microwave cavity to a kinetic energy of 3.7 MeV and focused through a holey mirror to a spot size of 200 µm FWHM in the interaction region of a gas phase experimental chamber. The pump pulses (50 µJ) are focused into the experimental chamber to a diameter of 250 µm FWHM and overlapped with the electron pulses at a 5° angle. The experimental response function including effects of the optical and electron pulse length as well as relative arrival time jitter is estimated to be 160 fs (see Supplementary Discussion 6). The sample 1,3-cyclohexadiene (CHD) is purchased from Sigma-Aldrich and used without further purification. We inject CHD vapor with a pulsed nozzle (100 µm orifice) into the interaction region of the experiment. Diffracted electrons are detected by a combination of a phosphor screen and an EMCCD camera. Based on the relative static and dynamic signal levels, we estimate that about 13 % of the molecules are excited (see Supplementary Discussion 7). We additionally perform pump pulse energy scans to confirm that we are in the linear absorption regime (see Supplementary Discussion 8). Time-dependent diffraction is measured at a series of delay time points between -1 and +1.8 ps in each scan. The separation between time delay points is 100 fs, except for the earliest and latest delay points, where it was considerably larger. At each time delay point, we integrate diffraction signal for 20 seconds. The full data set includes 166 such scans. The sequence of delay steps is randomized for every scan to avoid systematic errors. The camera images are azimuthally averaged and calibrated using a value of 0.0224 $Å^{-1}$/pixel based on diffraction of the molecule $CF_3I$. There is a center hole in the detector for transmitting the undiffracted electrons which cuts the s range at 0.5 $Å^{-1}$. Signal beyond 10.2 $Å^{-1}$ was not included in analysis due to limited signal-to-noise. The data evaluation is described in detail in Supplementary Discussion 2-5.

**Excited state wavepacket dynamics simulations:** *Ab-initio* multiple spawning (AIMS) wavepacket simulations[34] interfaced with GPU-accelerated[35,42–45] α-complete active space self-consistent field theory (α-CASSCF) are used to model the photodynamics of isolated CHD. The α-CASSCF method describes static correlation with a multireference CASSCF description of the electronic wavefunction,[40] while mimicking dynamic correlation effects through energy scaling (α).[35,41] Our active space consists of six electrons in four orbitals determined to minimize the average energy of the lowest two singlet states, within the 6-31G* basis set,[46] i.e. α-SA-2-CASSCF(6,4)/6-31G*. Electronic structure calculations are performed with TeraChem.[47–49] Following previous work,[35] we use an α value of 0.82. Electronic structure details and validation tests are given in Supplementary Discussion 9 and 10.

The first two singlet states ($S_0$ and $S_1$) are included in the dynamics. All required electronic structure quantities (energies, gradients, and nonadiabatic couplings) are calculated as needed with α-SA-2-CASSCF(6,4)/6-31G*. An adaptive timestep of 0.48 fs (20 au) (reduced to 0.12 fs (5 au) in regions with large nonadiabatic coupling) is used to propagate the centers of the trajectory basis functions (TBFs). A coupling threshold of 0.01 au (scalar product of nonadiabatic coupling and velocity vectors) demarcates spawning events generating new TBFs on different electronic states. Population transfer between TBFs is described by solving the time-dependent Schrödinger equation in the time-evolving TBF basis set. More details on AIMS are provided in Supplementary Discussion 11.

We simulate the first 1 ps of ultrafast dynamics for CHD by: 1) using AIMS to propagate the initial wavepacket for the first 500 fs or until all population has returned to the ground state, 2) stopping TBFs on the ground state when they are decoupled from other TBFs (off-diagonal elements of the Hamiltonian become small), and 3) adiabatically continuing these stopped TBFs using the positions and momenta from the last frame in AIMS as initial conditions for adiabatic molecular dynamics with unrestricted DFT using the Perdew-Burke-Ernzerof hybrid exchange-correlation functional,[50] i.e uPBE0/6-31G*. A total of 116 TBFs are propagated, with 86 of these being adiabatically continued on the ground state with DFT.

Following previous studies,[9,19–21] the computed time-dependent molecular diffraction from the AIMS/DFT trajectories are generated using the independent atom model (IAM) (see Supplementary Discussion 12). The computed diffraction signal is then processed in the same way as the experimental diffraction signal in order to generate simulated time-dependent PDFs.

Correspondence and requests for materials should be addressed to thomas.wolf@stanford.edu, wangxj@slac.stanford.edu, minitti@slac.stanford.edu, todd.martinez@stanford.edu.;

**References:**

1. Arruda, B. C. & Sension, R. J. Ultrafast polyene dynamics: the ring opening of 1,3-cyclohexadiene derivatives. *Phys. Chem. Chem. Phys.* **16,** 4439–4455 (2014).
2. Havinga, E. & Schlatmann, J. L. M. A. Remarks on the specificities of the photochemical and thermal transformations in the vitamin D field. *Tetrahedron* **16,** 146–152 (1961).
3. Woodward, R. B. & Hoffmann, R. The Conservation of Orbital Symmetry. *Angew. Chem. Int. Ed.* **8,** 781 (1969).
4. Bach, T. & Hehn, J. P. Photochemical Reactions as Key Steps in Natural Product Synthesis. *Angew. Chem. Int. Ed.* **50,** 1000–1045 (2011).
5. Irie, M. Diarylethenes for Memories and Switches. *Chem. Rev.* **100,** 1685–1716 (2000).
6. Deb, S. & Weber, P. M. The Ultrafast Pathway of Photon-Induced Electrocyclic Ring-Opening Reactions: The Case of 1,3-Cyclohexadiene. *Annu. Rev. Phys. Chem.* **62,** 19–39 (2011).

7. Hofmann, A. & de Vivie-Riedle, R. Quantum dynamics of photoexcited cyclohexadiene introducing reactive coordinates. *J. Chem. Phys.* **112,** 5054–5059 (2000).
8. Celani, P., Ottani, S., Olivucci, M., Bernardi, F. & Robb, M. A. What Happens during the Picosecond Lifetime of 2A1 Cyclohexa-1,3-diene? A CAS-SCF Study of the Cyclohexadiene/Hexatriene Photochemical Interconversion. *J. Am. Chem. Soc.* **116,** 10141–10151 (1994).
9. Ruan, C.-Y. *et al.* Ultrafast diffraction and structural dynamics: The nature of complex molecules far from equilibrium. *Proc. Natl. Acad. Sci.* **98,** 7117–7122 (2001).
10. Pullen, S. H., Anderson, N. A., Walker, L. A. & Sension, R. J. The ultrafast photochemical ring-opening reaction of 1,3-cyclohexadiene in cyclohexane. *J. Chem. Phys.* **108,** 556–563 (1998).
11. Cardoza, J. D., Dudek, R. C., Mawhorter, R. J. & Weber, P. M. Centering of ultrafast time-resolved pump–probe electron diffraction patterns. *Chem. Phys.* **299,** 307–312 (2004).
12. Attar, A. R. *et al.* Femtosecond x-ray spectroscopy of an electrocyclic ring-opening reaction. *Science* **356,** 54–59 (2017).
13. Adachi, S., Sato, M. & Suzuki, T. Direct Observation of Ground-State Product Formation in a 1,3-Cyclohexadiene Ring-Opening Reaction. *J. Phys. Chem. Lett.* **6,** 343–346 (2015).
14. Pemberton, C. C., Zhang, Y., Saita, K., Kirrander, A. & Weber, P. M. From the (1B) Spectroscopic State to the Photochemical Product of the Ultrafast Ring-Opening of 1,3-Cyclohexadiene: A Spectral Observation of the Complete Reaction Path. *J. Phys. Chem. A* **119,** 8832–8845 (2015).
15. Kotur, M., Weinacht, T., Pearson, B. J. & Matsika, S. Closed-loop learning control of isomerization using shaped ultrafast laser pulses in the deep ultraviolet. *J. Chem. Phys.* **130,** 134311 (2009).
16. Wolf, T. J. A. *et al.* Probing ultrafast ππ* / nπ* internal conversion in organic chromophores via K-edge resonant absorption. *Nat. Commun.* **8,** 29 (2017).
17. Stolow, A. & Underwood, J. G. Time-Resolved Photoelectron Spectroscopy of Nonadiabatic Dynamics in Polyatomic Molecules. in *Advances in Chemical Physics* **139,** 497 – 587 (Wiley, 2008).
18. Herbst, J., Heyne, K. & Diller, R. Femtosecond Infrared Spectroscopy of Bacteriorhodopsin Chromophore Isomerization. *Science* **297,** 822–825 (2002).

19. Ihee, H. *et al.* Direct Imaging of Transient Molecular Structures with Ultrafast Diffraction. *Science* **291,** 458–462 (2001).

20. Srinivasan, R., Lobastov, V. A., Ruan, C.-Y. & Zewail, A. H. Ultrafast Electron Diffraction (UED). *Helv. Chim. Acta* **86,** 1761–1799 (2003).

21. Minitti, M. P. *et al.* Imaging Molecular Motion: Femtosecond X-Ray Scattering of an Electrocyclic Chemical Reaction. *Phys. Rev. Lett.* **114,** 255501 (2015).

22. Dudek, R. C. & Weber, P. M. Ultrafast Diffraction Imaging of the Electrocyclic Ring-Opening Reaction of 1,3-Cyclohexadiene. *J. Phys. Chem. A* **105,** 4167–4171 (2001).

23. Küpper, J. *et al.* X-Ray Diffraction from Isolated and Strongly Aligned Gas-Phase Molecules with a Free-Electron Laser. *Phys. Rev. Lett.* **112,** 083002 (2014).

24. Jean-Ruel, H. *et al.* Ring-Closing Reaction in Diarylethene Captured by Femtosecond Electron Crystallography. *J. Phys. Chem. B* **117,** 15894–15902 (2013).

25. Ischenko, A. A., Weber, P. M. & Miller, R. J. D. Capturing Chemistry in Action with Electrons: Realization of Atomically Resolved Reaction Dynamics. *Chem. Rev.* (2017). doi:10.1021/acs.chemrev.6b00770

26. Gao, M. *et al.* Mapping molecular motions leading to charge delocalization with ultrabright electrons. *Nature* **496,** 343–346 (2013).

27. Zimmerman, H. E. & Nesterov, E. E. Development of Experimental and Theoretical Crystal Lattice Organic Photochemistry: The Quantitative Cavity. Mechanistic and Exploratory Organic Photochemistry [1]. *Acc. Chem. Res.* **35,** 77–85 (2002).

28. Zimmerman, H. E. & Zuraw, M. J. Photochemistry in a box. Photochemical reactions of molecules entrapped in crystal lattices: mechanistic and exploratory organic photochemistry. *J. Am. Chem. Soc.* **111,** 7974–7989 (1989).

29. Yang, J. *et al.* Diffractive imaging of a rotational wavepacket in nitrogen molecules with femtosecond megaelectronvolt electron pulses. *Nat. Commun.* **7,** 11232 (2016).

30. Yang, J. *et al.* Diffractive Imaging of Coherent Nuclear Motion in Isolated Molecules. *Phys. Rev. Lett.* **117,** 153002 (2016).

31. Yang, J. *et al.* Imaging CF3I conical intersection and photodissociation dynamics by ultrafast electron diffraction. *Science* accepted (2018).

32. Kirrander, A. & Weber, P. M. Fundamental Limits on Spatial Resolution in Ultrafast X-ray Diffraction. *Appl. Sci.* **7,** 534 (2017).

33. Shao, J. & Tu, D. *The Jackknife and Bootstrap*. (Springer-Verlag, 1995).

34. Ben-Nun, M., Quenneville, J. & Martínez, T. J. Ab Initio Multiple Spawning: Photochemistry from First Principles Quantum Molecular Dynamics. *J. Phys. Chem. A* **104,** 5161–5175 (2000).

35. Snyder, J. W., Parrish, R. M. & Martínez, T. J. α-CASSCF: An Efficient, Empirical Correction for SA-CASSCF To Closely Approximate MS-CASPT2 Potential Energy Surfaces. *J. Phys. Chem. Lett.* **8,** 2432–2437 (2017).

36. Kosma, K., Trushin, S. A., Fuss, W. & Schmid, W. E. Cyclohexadiene ring opening observed with 13 fs resolution: coherent oscillations confirm the reaction path. *Phys. Chem. Chem. Phys.* **11,** 172–181 (2009).

37. Kuthirummal, N., Rudakov, F. M., Evans, C. L. & Weber, P. M. Spectroscopy and femtosecond dynamics of the ring opening reaction of 1,3-cyclohexadiene. *J. Chem. Phys.* **125,** 133307 (2006).

38. Harris, D. A., Orozco, M. B. & Sension, R. J. Solvent Dependent Conformational Relaxation of cis-1,3,5-Hexatriene. *J. Phys. Chem. A* **110,** 9325–9333 (2006).

39. Weathersby, S. P. *et al.* Mega-electron-volt ultrafast electron diffraction at SLAC National Accelerator Laboratory. *Rev. Sci. Instrum.* **86,** 073702 (2015).

40. Roos, B. O. The Complete Active Space Self-Consistent Field Method and Its Applications in Electronic Structure Calculations. *Adv. Chem. Phys.* **69,** 399–445 (1987).

41. Frutos, L., Andruniow, T., Santoro, F., Ferre, N. & Olivucci, M. Tracking the excited-state time evolution of the visual pigment with multiconfigurational quantum chemistry. *Proc. Natl. Acad. Sci. U. S. A.* **104,** 7764 (2007).

42. Hohenstein, E. G., Luehr, N., Ufimtsev, I. S. & Martínez, T. J. An atomic orbital-based formulation of the complete active space self-consistent field method on graphical processing units. *J. Chem. Phys.* **142,** 224103 (2015).

43. Snyder, J. W., Hohenstein, E. G., Luehr, N. & Martínez, T. J. An atomic orbital-based formulation of analytical gradients and nonadiabatic coupling vector elements for the state-averaged complete active space self-consistent field method on graphical processing units. *J. Chem. Phys.* **143,** 154107 (2015).

44. Snyder, J. W., Fales, B. S., Hohenstein, E. G., Levine, B. G. & Martínez, T. J. A direct-compatible formulation of the coupled perturbed complete active space self-consistent field equations on graphical processing units. *J. Chem. Phys.* **146,** 174113 (2017).

45. Snyder, J. W., Curchod, B. F. E. & Martínez, T. J. GPU-Accelerated State-Averaged Complete Active Space Self-Consistent Field Interfaced with Ab Initio Multiple Spawning Unravels the Photodynamics of Provitamin D3. *J. Phys. Chem. Lett.* **7,** 2444–2449 (2016).

46. Hehre, W. J., Ditchfield, R. & Pople, J. A. Self—Consistent Molecular Orbital Methods. XII. Further Extensions of Gaussian—Type Basis Sets for Use in Molecular Orbital Studies of Organic Molecules. *J Chem Phys* **56,** 2257–2261 (1972).

47. Ufimtsev, I. S. & Martínez, T. J. Quantum Chemistry on Graphical Processing Units. 1. Strategies for Two-Electron Integral Evaluation. *J. Chem. Theory Comput.* **4,** 222–231 (2008).

48. Ufimtsev, I. S. & Martinez, T. J. Quantum Chemistry on Graphical Processing Units. 2. Direct Self-Consistent-Field Implementation. *J. Chem. Theory Comput.* **5,** 1004–1015 (2009).

49. Ufimtsev, I. S. & Martinez, T. J. Quantum Chemistry on Graphical Processing Units. 3. Analytical Energy Gradients, Geometry Optimization, and First Principles Molecular Dynamics. *J. Chem. Theory Comput.* **5,** 2619–2628 (2009).

50. Perdew, J. P. Density-functional approximation for the correlation energy of the inhomogeneous electron gas. *Phys Rev B* **33,** 8822–8824 (1986).

**Acknowledgements:**

This work was supported by the U.S. Department of Energy, Office of Science, Basic Energy Sciences, Chemical Sciences, Geosciences, and Biosciences Division. The experimental part of this research was performed at the SLAC megaelectronvolt ultrafast electron diffraction facility, which is supported in part by the DOE BES SUF Division Accelerator & Detector R&D program, the Linac Coherent Light Source (LCLS) Facility, and SLAC under contract Nos. DE-AC02-05-CH11231 and DE-AC02-76SF00515. M. G. is funded via a Lichtenberg Professorship of the Volkswagen Foundation. D. M. S. is grateful to the NSF for a graduate fellowship. J. P. F. N acknowledges the support by the Wild Overseas Scholars Fund of Department of Chemistry, University of York. K. W. and M. C. are supported by the U.S. Department of Energy Office of Science, Basic Energy Sciences under Award No. DE-SC0014170. PMW is supported by the U.S. Department of Energy, Office of Science, Basic Energy Sciences, under Award No. DE-SC0017995. A. K. is supported by the Carnegie Trust for the Universities of Scotland (grant ref. CRG050414) and an RSE/Scottish Government Sabbatical Research Grant (ref. 58507).

# Supplementary Information

# Imaging the Photochemical Ring-Opening of 1,3-Cyclohexadiene by Ultrafast Electron Diffraction

T. J. A. Wolf[1,*], D. M. Sanchez[1,2], J. Yang[1,3], R. M. Parrish[1,2], J. P. F. Nunes[4,5], M. Centurion[5], R. Coffee[3], J. P. Cryan[1], M. Gühr[1,6], K. Hegazy[1,7], A. Kirrander[8], R. K. Li[3], J. Ruddock[9], X. Shen[3], T. Veccione[3], S. P. Weathersby[3], P. M. Weber[9], K. Wilkin[5], H. Yong[9], Q. Zheng[3], X. J. Wang[3,*], M. P. Minitti[3,*], T. J. Martínez[1,2,*]

[1]Stanford PULSE Institute, SLAC National Accelerator Laboratory, Menlo Park, USA.

[2]Department of Chemistry, Stanford University, Stanford, USA.

[3]SLAC National Accelerator Laboratory, Menlo Park, USA.

[4]Department of Chemistry, University of York, Heslington, York, UK.

[5]Department of Physics and Astronomy, University of Nebraska-Lincoln, Lincoln, USA.

[6]Institut für Physik und Astronomie, Universität Potsdam, Potsdam, Germany.

[7]Department of Physics, Stanford University, Stanford, USA.

[8]EaStCHEM, School of Chemistry, University of Edinburgh, Edinburgh EH9 3FJ, United Kingdom.

[9]Department of Chemistry, Brown University, Providence, USA.

## Table of Contents

## Supplementary Discussion

### 1. Stereochemistry of 1,3-cyclohexadiene

The ground state minimum of 1,3-cyclohexadiene (CHD) exhibits $C_2$ symmetry, which makes the molecule chiral. The barrier for ground state isomerization between enantiomers is, however, very low.[1] Therefore, samples containing only one of the two enantiomers have never been generated. Two of the photoproduct isomers, the cZc and cZt minima of 1,3,5-hexatriene (HT) are also chiral. Thus, there exist two enantiomer minima for each of those isomers on the ground state potential energy surface. However, since in the present experiment we are exciting a racemic mixture of the two 1,3-cyclohexadiene enantiomers, no stereochemical information can be obtained from the diffraction signal. We therefore do not mention the presence of additional enantiomer minima in the main text.

### 2. Simulation of steady state molecular diffraction for a fixed classical molecular geometry

Molecular diffraction $I_{Mol}(s)$ is simulated within the independent atom model (IAM) from a molecular geometry using

$$I_{Mol}(s) = \sum_i \sum_{i \neq j} |f_i(s)| \cdot |f_j(s)| \cos(\eta_i - \eta_j) \frac{\sin(s \cdot R_{ij})}{s \cdot R_{ij}}$$

The summation is over all atomic distances $R_{ij}$ of the geometry. The elastic scattering factors of atoms i and j, $f_{i,j}(s)$, are calculated using the ELSEPA program,[2] and $\eta_{i,j}$ are scattering phases. $I_{Mol}(s)$ is used to compute pair distribution functions (see below). Ground state minima and saddle point geometries of CHD and HT (see Figs. 1 and 2 of the main text) are optimized using

MP2/cc-pVDZ as implemented in Firefly.[3] Single-point energies at the EOM-CCSD/aug-cc-pVDZ level of theory using the GAMESS program package are shown in Fig. 1 of the main text.[4]

### 3. Determination of molecular diffraction from the experimental data

The radially integrated signal $\left(I_{Tot}(s)\right)$ contains contributions from molecular diffraction $\left(I_{Mol}(s)\right)$, scattering from each atom in the molecule $\left(I_{At}(s)\right)$, and additional background contributions, $\left(I_{BKG}(s)\right)$, from non-uniform detector response, gas scattering, and laser light:

$$I_{Tot}(s) = I_{At}(s) + I_{Mol}(s) + I_{BKG}(s)$$

The structural information about the molecule is contained in $I_{Mol}(s)$. To extract the $I_{Mol}(s)$ contribution in the static diffraction signal before time zero, we compare it to a simulation (see above) based on an optimized geometry of CHD. We use the zero points of the simulation to determine where the contribution of $I_{Mol}(s)$ to $I_{Tot}(s)$ is zero and fit a weighted exponential function through the experimental intensities at these s-values. Subtracting the fit from the experimental signal efficiently removes both $I_{At}(s)$ and $I_{BKG}(s)$ while leaving $I_{Mol}(s)$ in a wide range of s values. We emphasize that we use this method for extracting $I_{Mol}(s)$ only for the static PDF in Fig. 2a of the main text. It is not required for ΔPDFs (see below), since the subtraction of static diffraction efficiently removes the non- $\Delta I_{Mol}(s)$ background.

It is convenient to express the structural information in terms of the modified diffraction intensity:

$$sM(s) = s\frac{I_{Mol}(s)}{I_{At}(s)}$$

where

$$I_{At}(s) = \sum_i \left| f_i(s) \right|^2$$

## 4. Determination of static pair distribution functions (PDF)

The sinusoidal transformation from momentum transfer space into real space yields an atomic pair distribution function (PDF)

$$PDF(R) = \int_0^{s_{Max}} sM(s)\sin(sR)e^{-ks^2}\, ds$$

The increasing noise at high s values is damped using a Gaussian function with k = 0.05 Å$^2$. We observe additional background contributions to the $I_{Mol}(s)$ signal for 0.5 Å$^{-1}$ < s < 1.3 Å$^{-1}$ from the main electron beam, which are not removed by the fit subtraction. Omitting this s range together with the range 0 Å$^{-1}$ < s < 0.5 Å$^{-1}$ induces unphysical artifacts in the PDF, e.g. considerable negative PDF amplitudes outside the bond distances of CHD. We, therefore, fill the experimental signal in the range 0 Å$^{-1}$ < s < 1.3 Å$^{-1}$ with scaled values from the simulation. The procedure removes the artifacts but otherwise does not considerably change the PDF shape. Again, this method is only used for the static PDF in Fig. 2a of the main text.

## 5. Determination of difference pair distribution functions

Other than in the case of the static, $I_{Mol}(s)$, we do not have to rely on subtracting a fitted background from the experimental $I_{Tot}(s)$ to obtain the signatures of pump-induced structural changes in the molecules $\Delta I_{Mol}(s)$. We simply subtract the static signal $I_{Tot}(s)$ before time zero from all the time steps $I_{Tot}(s,t)$. Scaling by s and $I_{At}(s)$ results in the modified difference diffraction intensity

$$\Delta sM(s,t) = s\frac{I_{Mol}(s,t) - I_{Mol}(s)}{I_{At}(s)}$$

We, however, observe a time-dependent contribution in the range 0.5 $Å^{-1}$ < s < 1.3 $Å^{-1}$, where the amplitude of simulated $\Delta sM(s,t)$ from our wavepacket calculations (see below) is virtually zero. We speculate that the offset originates from the focus of the pump pulse, which occurred downstream of the interaction region. It is possible that we created a plasma of remaining sample gas there, which distorted the profile of the undiffracted beam. We correct for the artifact by subtracting an exponential decay from the difference diffraction that sets it to zero at 0.5 $Å^{-1}$ but does not affect the signal at s values beyond 1.3 $Å^{-1}$. Accordingly, we set the amplitudes in the range between 0 $Å^{-1}$ and 0.5 $Å^{-1}$, where we do not have experimental signal, to zero. We obtain difference pair distribution functions $(\Delta PDF(R,t))$ from:

$$\Delta PDF(R,t) = \int_{0}^{s_{Max}} \Delta sM(s,t)\sin(sR)e^{-ks^2}\, ds$$

To ensure that the $\Delta PDFs$ were not biased by data treatment of the diffraction below 1.5 $Å^{-1}$, we compared them to $\Delta PDFs$, where the section was replaced by zeros. This leads to distortions of the $PDFs$ (see Supplementary Figure 7) but does not alter our findings about the dynamics.

The center-of-mass curves in Fig. 3c and 3d of the main text are calculated from the intensities in peak γ area according to

$$COM(t)=\frac{\sum_i R_i PDF(R_i,t)}{\sum_i R_i}$$

The error bars are obtained by bootstrapping from $PDF(R_i,t)$.

## 6. Determination of the experimental instrument response function

In contrast to the electronic response probed by time-resolved electronic spectroscopy, nuclear response to photoexcitation is not quasi-instantaneous on the fs timescale. Hence, the onset of any time-dependent diffraction signal at time zero represents a convolution of the instrument response function and the molecular response function and neither the exact time zero nor the instrument response function is easily determined. Therefore, we estimate the instrument response function by fitting the AIMS-simulated $\Delta PDFs$ with an offset correction of time zero and convoluted with a Gaussian response function to the experimental $\Delta PDFs$. The parameters of the fit are the time zero offset, the width of the response function, and an overall scaling factor. The fit results in a Gaussian response function with a FWHM of 160 ± 10 fs.

## 7. Estimation of the excitation ratio and intensity scan

We estimate the excitation ratio as the ratio between absolute transient diffraction $\left(A_{T,Exp}\right)$ and absolute static diffraction $\left(A_{S,Exp}\right)$. $\left(A_{T,Exp}\right)$ is averaged between 0.6 ps and 0.8 ps and integrated over an area between 2 Å$^{-1}$ and 8 Å$^{-1}$. $\left(A_{S,Exp}\right)$is integrated over the same momentum transfer range. We correct this ratio by the estimated ratio in per-molecule transient and static signal based on the first step (static signal) and averaged steps between 0.6 and 0.8 ps from the AIMS simulations $\left(A_{S,Sim}, A_{T,Sim}\right)$:

$$f = \frac{A_{T,Exp}}{A_{S,Exp}} \cdot \frac{A_{S,Sim}}{A_{T,Sim}}$$

## 8. Pump pulse energy scan

We obtained difference diffraction patterns at five different pump pulse intensities. In Supplementary Figure 8, we show their pump pulse energy-dependent integrated absolute intensity in a momentum transfer range between 2.5 $Å^{-1}$ and 7.5 $Å^{-1}$. The graph shows a clear saturation effect above 100 µJ pump pulse energy. For our time-dependent experiments, we kept the pump pulse energy at 50 % of the saturation value.

## 9. Electronic structure validation

We employ *ab-initio* multiple spawning (AIMS) wavepacket simulations[5-8] interfaced with GPU-accelerated α-complete active space self-consistent field theory (α-CASSCF)[9-13] to study the ultrafast dynamics of 1,3 cyclohexadiene (CHD) in the gas phase. α-CASSCF is an empirical correction to state-averaged CASSCF that scales the state-specific energy splittings by a constant, α, while leaving the state-averaged energy unaffected. The use of a scaling factor roughly incorporates the effects of dynamic electron correlation. Although features of the α-CASSCF potential energy surface (PES) (e.g. critical point positions, relative energies, and barrier heights) will usually differ from SA-CASSCF, the location of the seam space for conical intersections (CI) remains unchanged and the topology of the potential energy surface around intersections exhibits the correct branching space dimensionality. This is of particular importance for CHD as its ultrafast dynamics has been shown to include a nonradiative relaxation pathway through a $S_0/S_1$ CI.[14-18] All electronic structure calculations (i.e. energies, gradients, and nonadiabatic couple vectors (NACV)) are performed with the TeraChem electronic structure package.[19-21]

In this work, we use an active space consisting of six electrons in four orbitals determined to minimize the average energy of the lowest two singlet states, in conjunction with the 6-31G* basis set, i.e. α-SA-2-CASSCF(6,4)/6-31G*. Previous SA-3-CASSCF(6,4)/6-31G* calculations[16] show that due to the topography of the $S_2/S_1$ CI, it plays only a minor role in the nonadiabatic dynamics of CHD. While propagating on the $S_1$ electronic state, the nuclear wavepacket largely avoids the $S_2/S_1$ CI, resulting in minimal population transfer to the $S_2$ state. Therefore, only the two lowest singlet states ($S_0$ and $S_1$) are included in the α-SA-2-CASSCF(6,4) simulations.

The often-used active space consisting of six electrons in six orbitals (two σ/σ* orbitals and four π/π* orbitals) incorrectly places the optically bright $^1B$ state above the optically dark $^2A$ state.[16] This is due to the different rates at which dynamic and static electron correlation are recovered as the active space increases. Our chosen active space with six electrons in four orbitals provides a more balanced treatment of static and dynamic electron correlation effects. Critical points (Frank-Condon point (FC), $S_1/S_0$-closed minimum energy conical intersection (MECI), $S_1/S_0$-open MECI, and hexatriene (HT) cis-Z-cis (cZc-HT)) along the CHD ring-opening pathway are computed with α-SA-2-CASSCF(6,4)/6-31G* using the DL-FIND optimization package[22] (Supplementary Figure 9). These show excellent agreement with previous multistate complete active space 2$^{nd}$ order perturbation theory (MS-CASPT2) calculations using a (6,6) active space.[16] In addition, $S_0$ and $S_1$ energies, active space molecular orbitals (MO), and CI eigenvectors for these critical points are shown in Supplementary Figure 10 and Supplementary Table 1, respectively.

## 10. Preparation of initial conditions for the AIMS simulations

An ultraviolet (UV) electronic absorption spectrum (see Supplementary Figure 11) is generated from 500 geometries sampled from a harmonic Wigner distribution corresponding to the PBE0/6-31G* ground state optimized structure. Single point energy calculations are performed at the α-SA-2-CASSCF(6,4)/6-31G* level for all 500 initial conditions and their $S_0 \rightarrow S_1$ excitation energies homogeneously broadened using Gaussian functions with a full width half maximum (FWHM) of 0.2 eV. Positions and momenta for 30 different initial conditions selected from the 500 phase space points used to generate the electronic absorption spectrum are used to initiate the AIMS dynamics. These initial conditions are selected under the constraint that their $S_0 \rightarrow S_1$ transition energy was within 0.3 eV of the pump pulse (4.65 eV) used in the experiment after applying a 0.22 eV red-shift of the theoretical spectrum to align the theoretical and experimental absorption maxima. These initial conditions are then placed on the $S_1$ surface and propagated with AIMS.

### 11. AIMS dynamics

Ab initio multiple spawning (AIMS) is a nonadiabatic dynamics method aimed at describing photodynamical processes involving multiple electronic states. In the following, we present a brief introduction to the working equations of AIMS and direct the reader elsewhere for a more complete description.[6,23] Using the Born-Huang representation,[24] the exact molecular wavefunction can be separated into electronic and nuclear contributions:

$$\Psi\left(\mathbf{r},\mathbf{R},t\right)=\sum_{I}\chi_{I}\left(\mathbf{R},t\right)\phi_{I}\left(\mathbf{r};\mathbf{R}\right)$$

where $\chi_I\left(\mathbf{R},t\right)$ denotes the time-dependent nuclear wavefunction associated with electronic state I and $\phi_I\left(\mathbf{r};\mathbf{R}\right)$ is the electronic wavefunction for state I at nuclear configuration **R**. In the adiabatic representation, $\phi_I(\mathbf{r};\mathbf{R})$ is expanded into an orthonormal electronic basis consisting of eigenfunctions of the time-independent electronic Schrödinger equation (TIESE) parametrically dependent on nuclear configuration **R**. Under the AIMS ansatz, $\chi_I(\mathbf{R},t)$ is represented as a superposition of frozen Gaussian functions called trajectory basis functions (TBF):

$$\chi_I(\mathbf{R},t) = \sum_{k=1}^{N_I(t)} c_k^I(t)\chi_k^I\left(\mathbf{R};\bar{\mathbf{R}}_k^I(t),\bar{\mathbf{P}}_k^I(t),\bar{\gamma}_k^I(t),\alpha_k^I\right)$$

where $N_I(t)$ represents the total number of TBFs on electronic state I, $c_k^I(t)$ is the time-dependent complex coefficient of the *k*th TBF, $\alpha_k^I$ is the frozen TBF width, and $\chi_k^I(...)$ is a multidimensional frozen Gaussian that is expressed as a product of one-dimensional Gaussian functions corresponding to the 3N nuclear degrees of freedom:

$$\chi_k^I\left(\mathbf{R};\bar{\mathbf{R}}_k^I(t),\bar{\mathbf{P}}_k^I(t),\bar{\gamma}_k^I(t),\alpha_k^I\right) = e^{i\bar{\gamma}_k^I(t)t}\prod_{\rho=1}^{3N}\chi_{\rho_k}^I\left(R;\bar{R}_{\rho_k}^I(t),\bar{P}_{\rho_k}^I(t),\alpha_{\rho_k}^I\right)$$

where $\chi_{\rho_k}^I(...)$ is:

$$\chi_{\rho_k}^I\left(R;\bar{R}_{\rho_k}^I(t),\bar{P}_{\rho_k}^I(t),\alpha_{\rho_k}^I\right) = \left(\frac{2\alpha_{\rho_k}^I}{\pi}\right)^{1/4}\prod_{\rho=1}^{3N}\exp\left[-\alpha_{\rho_k}^I\left(R_{\rho_k} - \bar{R}_{\rho_k}^I(t)\right)^2 + i\bar{P}_{\rho_k}^I(t)\left(R_{\rho_k} - \bar{R}_{\rho_k}^I(t)\right)\right]$$

In AIMS, each TBF evolves adiabatically along one Born-Oppenheimer electronic surface. The time-dependent positions and momenta, $\left(\bar{R}_{\rho_k}^I(t),\bar{P}_{\rho_k}^I(t)\right)$, of the TBFs are propagated classically according to Hamilton's equations of motion on the given electronic state:

$$\frac{\partial\bar{\mathrm{R}}_{\rho_k}^I(t)}{\partial t} = \frac{\bar{\mathrm{P}}_{\rho_k}^I(t)}{m_\rho}$$

$$\frac{\partial \bar{P}^{I}_{\rho_k}(t)}{\partial t} = -\left.\frac{\partial E_I(\mathbf{R})}{\partial R_{\rho_k}}\right|_{\bar{R}^{I}_{\rho_k}(t)}$$

where $m_\rho$ is the mass for the $\rho$th nuclear coordinate and $E_I(\mathbf{R})$ is the electronic energy of state I with nuclear configuration **R**. The nuclear phase, $\bar{\gamma}^{I}_{k}(t)$, is propagated semiclassically according to the classical Lagrangian:

$$\frac{\partial \bar{\gamma}^{I}_{k}}{\partial t} = \sum_{\rho=1}^{3N} \frac{\left(\bar{P}^{I}_{\rho_k}(t)\right)^2}{2m_\rho} - E_I\left(\bar{\mathbf{R}}^{I}_{k}(t)\right)$$

The evolution of the time-dependent amplitudes is governed by the time-dependent Schrodinger equation (TDSE) written in matrix form as:

$$\frac{d\mathbf{C}^{I}(t)}{dt} = -i\left(\mathbf{S}^{-1}_{II}\right)\left\{\left[\mathbf{H}_{II} - i\dot{\mathbf{S}}_{II}\right]\mathbf{C}^{I} + \sum_{J\neq I}\mathbf{H}_{IJ}\mathbf{C}^{J}\right\}$$

where **S** and $\dot{\mathbf{S}}$ are the nuclear overlap matrix and its right-acting time-derivative, respectively, and **H** is the Hamiltonian. The matrix elements of the Hamiltonian are computed using the zeroth-order saddle-point (SP) approximation, which involves evaluating the zeroth-order Taylor expansion of the PES and/or NACV around the centroid position between each pair of TBFs.[23] In addition to the SP approximation, the independent first generation (IFG) approximation is used to describe the initial nuclear wavepacket at time t = 0 as a swarm of independent initial TBFs each with their own positions and momenta sampled from a harmonic Wigner distribution, i.e. initial conditions (IC). The IFG is justified by the assumption that the initial nuclear wavepacket spreads quickly in phase space and that the TBFs will soon become uncoupled and evolve independently.[23] Unlike the initial TBFs, the spawned TBFs from each IC remain coupled during the course of the dynamics and their separation naturally accounts for decoherence of the nuclear wavefunction on multiple electronic states.

As a direct consequence of the IFG, the population of the nuclear wavepacket on a given electronic state can be computed from an incoherent average of the population results for each of the 30 ICs. Within a set of TBFs arising from the same IC, they should be computed accounting for all coherences between TBFs. Therefore, the total population on the *I*th electronic state for a set of $N_{IC}$ initial conditions is computed as:[25]

$$P_I(t) = \frac{1}{N_{IC}} \sum_{M=1}^{N_{IC}} \sum_{k,l}^{N_I^M(t)} \left(c_{k,M}^I(t)\right)^* S_{kl,M}^{II}(t) c_{l,M}^I(t)$$

where $N_{IC}$ is the total number of initial conditions, $N_I^M(t)$ is the total number of TBFs on the *I*th electronic state for the *M*th initial condition, $c_{k,M}^I(t)$ is the amplitude for the *k*th TBF (on the *I*th electronic state) for the *M*th initial condition, and $S_{kl,M}^{II}(t)$ is the overlap matrix element between TBFs *k* and *l* (both on state *I*) for the *M*th initial condition. The population of the nuclear wavepacket ($S_1$ red; $S_0$ black) is shown in Supplementary Figure 12, along with the contributions from individual ICs (grey). CHD exhibits an incubation period where essentially no population is transferred to the ground state for ~20 fs. The population on the $S_1$ state decays quickly, with 50% population transferred to the $S_0$ state within 100 fs. Nearly all of the population has returned to the ground state within 400 fs. The $S_1$ population decay curve is fitted to a monoexponential curve resulting in a computed lifetime of 139 ± 25fs, which is in line with previous studies.[16-18,26] However, this disagrees with previous studies predicting shorter excited state lifetimes (76-106fs) using linear response time-dependent density functional theory (LR-TDDFT) coupled to Tully's Surface Hopping (TSH).[27,28] This discrepancy may be due to the inability of LR-TDDFT to correctly describe the correct 3$n$-8 dimensionality (where $n$ is the number of atoms) of the $S_0/S_1$ CI and the double excitation character of the $S_1$ adiabat near the CI, leading to very short excited state lifetimes.[29]

The influence of the basis set size is examined by comparing the potential energies computed with both 6-31G* and 6-31G** basis sets along the trajectory of one TBF propagating on $S_1$ (Supplementary Figure 13). There is a slight systematic shift in energy of 0.35 eV, but the computed potential energies are very similar (and nearly parallel) between the two basis sets. Therefore, the 6-31G* basis set was chosen for all dynamics.

## 12. Generation of time-dependent diffraction based on the wavepacket simulations

The modified difference diffraction signal defined above, $\Delta sM\left(s,t\right)$, is generated from the AIMS/DFT trajectories using the IAM and converted to $\Delta PDF\left(R,t\right)$ using identical code and procedures as for the experimental data. The total diffraction signal, $I_{Mol}\left(s,t\right)$, is computed as an average over all 30 ICs, where the diffraction signal for a specific IC is approximated as an incoherent sum over weighted diffraction signals from individual TBFs:

$$I_{Mol}\left(s,t\right)=\frac{1}{N_{IC}}\sum_{M=1}^{N_{IC}}\sum_{k}^{N_{TBF}^{M}(t)}n_k^M\left(t\right)I_{Mol}^{k,M}\left(s,t\right)$$

where $N_{IC}$ is the number of ICs, $N_{TBF}^{M}\left(t\right)$ is the number of TBFs at time $t$ for the $M$th IC, $n_k^M\left(t\right)$ and $I_{Mol}^{k,M}\left(s,t\right)$ are the weight and diffraction signal for the $k$th TBF of the $M$th IC at time $t$, respectively. The expression for $I_{Mol}^{k,M}\left(s,t\right)$ is identical to that used in the experimental diffraction signal, augmented with a Gaussian factor to account for the finite width of the TBFs:[30]

$$I_{Mol}^{k,M}(s,t)=\sum_i\sum_{j\neq i}\left|f_i(s)\right|\cdot\left|f_j(s)\right|\cos\left(\eta_i-\eta_j\right)\frac{\sin\left(s\cdot R_{ij}(t)\right)}{s\cdot R_{ij}(t)}e^{-\left(\alpha_i^2+\alpha_j^2\right)s^2}$$

where $\alpha_i$ and $\alpha_j$ represent the finite widths for the atoms used in the TBFs.[31] These widths are taken to be element specific and are 0.112 Å/0.249 Å for carbon/hydrogen. $R_{ij}$ is the interatomic distance between the *i*th and *j*th atoms taken from the centroids of the AIMS/DFT TBFs. The weight $n_k^M(t)$ is evaluated according to the bra-ket averaged Taylor expansion (BAT) method:[32]

$$n_k^M(t)=\frac{1}{2}\sum_l^{N_I^M(t)}\left[c_k^*(t)S_{kl}c_l(t)+c_l^*(t)S_{lk}c_k(t)\right]$$

It is important to note, that the complex amplitudes are time-independent during the DFT adiabatic dynamics and are held constant at the value from the last frame of their corresponding AIMS trajectory. This is valid because the ground state TBFs are effectively uncoupled from all other TBFs.

In order to account for the adaptive timesteps, we rebin the AIMS and DFT trajectories onto a time grid with common, constant 2 fs step size. The diffraction contributions are averaged over the initial conditions. The theoretical $\Delta sM(s,t)$ analog to the experimental diffraction signal is computed by subtracting the static diffraction of CHD from all time-bins. In the case of the simulations, this is the diffraction of the initial conditions, $\Delta sM(s,0)$. As was also performed for the experimental diffraction data, we set values < 0.5 $Å^{-1}$ to zero and use the same parameters for real space transformation. As shown in Supplementary Figure 6, top left and right, setting the sM(s< 0.5 $Å^{-1}$) to zero introduces only minor changes to the $\Delta PDFs$. The resulting $\Delta PDFs$ are convolved with a 160 fs temporal Gaussian to match the experimental instrument response function (see Supplementary Figure 6, bottom). The error bars in Fig. 3 of the main text represent the standard deviation of the ΔPDF values as evaluated by bootstrap sampling from the initial conditions. They are a measure for the level of convergence of the simulation for the employed number of ICs. Our simulations show that (46 ± 7) % of the excited state population relaxes to the ground state without undergoing ring-opening, which is in agreement with previous work.[33] However, the contribution of this part of the population to the *ΔPDFs* is negligible, since structural changes are considerably smaller.

### 13. Classifying Hexatriene Isomers from Structural Dynamics on $S_0$

The time-dependent center of mass in Figure 3c/d strongly suggests coherent oscillation of population between the cZc-HT, cZt-HT, and tZt-HT isomers. This was further investigated by binning geometries along ground-state TBFs into one of the four CHD photoproducts (CHD, cZc-HT, cZt-HT, and tZt-HT). Snapshots taken every 5fs along all 86 ground state TBFs were optimized at the uPBE0/6-31G* level of theory and binned into one of the four isomers based on dihedral angles ($\Phi_1$, $\Phi_2$, and $\Phi_3$) and $R_1$ of the optimized structure. For details on the binning criteria, see Supplementary Figure 14. Due to the ground state TBFs being sufficiently uncoupled from all other TBFs, the BAT expression shown above reduces to:

$$n_k^M(t) = \left| c_k^M(t) \right|^2$$

where the overlap elements between TBFs vanish and the total population of a specific isomer $L$ at time t on the ground-state, $P_L(t)$ is computed by:

$$P_L(t) = \frac{1}{N_{IC}} \sum_{M=1}^{N_{IC}} \left[ \frac{\sum_k^{N_{TBF}^M(t)} c_{k,M}^*(t)\, c_{k,M}(t)\, \delta\left(L, I\left(\bar{R}_{k,M}(t)\right)\right)}{\sum_k^{N_{TBF}^M(t)} c_{k,M}^*(t)\, c_{k,M}(t)} \right]$$

where $L$ is defined as one of CHD, cZc-HT, cZt-HT, or tZt-HT, $c_{k,M}(t)$ is the amplitude of the $k$th TBF and $M$th IC, $N_{IC}$ is the total number of initial conditions, and $\delta(\rightleftharpoons)$ is a Kronecker-delta function, and $I(R)$ represents the isomer classification of the geometry given by $R$.

The population of Hexatriene isomers on the ground-state at time $t$ after photoexcitation, along with the total $S_0$ population, is shown in Supplementary Figure 15. In the first 100fs after photoexcitation, nearly all ground state population is either CHD or cZc-HT. At 200fs, the cZc-HT population is converted entirely into cZt-HT and tZt-HT, corresponding to an increase in the time-dependent center-of-mass in peak γ. At 400fs, nearly all tZt-HT population is converted back into cZc-HT and cZt-HT. Lastly, we observe a revival of the cZc-HT and cZt-HT signal as nearly all tZt-HT is rapidly converted. The low frequency, breathing-like motion of the time-dependent center-of-mass is highly correlated with the coherent exchange of population between cZc-HT and tZt-HT on the ground state. In addition, the observed branching ratio between CHD and HT is approximately 1:1, which agrees well with previous CASSCF studies.

A complementary view of the population of Hexatriene isomers on the ground state after quenching to $S_0$ is shown in Supplementary Figure 16. Here, the ground-state TBFs have been shifted so that $t$=0 corresponds to the time when they were spawned from $S_1$. As in Supplementary Figure 15, all ground-state TBFs start on the ground-state as either CHD or cZc-HT. The CHD population stays roughly constant making up ~50% of the wavefunction, implying that there is negligible conversion from CHD to a ring-opened isomer on the ground state within the first 1ps after photoexcitation. Approximately 90fs after quenching to $S_0$, cZc-HT TBFs convert almost entirely into cZt-HT TBFs and then into tZt-HT, with the largest probability of a TBF belonging to tZt-HT occurring at ~150fs after its birth on $S_0$. During the remainder of the simulation, the ring-opened $S_0$ TBFs continue to convert between isomers. Also worth noting, in both Supplementary Figures 15 and 16, a steady increase of cZc-HT population is observed as population oscillates between cZc-HT and tZt-HT, trapping cZt-HT with each cycle. We note that the lowest energy ring-opened isomer is tZt-HT and this is not the dominant isomer for most of the time shown. Thus, the dynamics of isomer interconversion is not well-predicted by a statistical theory on this time scale.

## Supplementary Figures

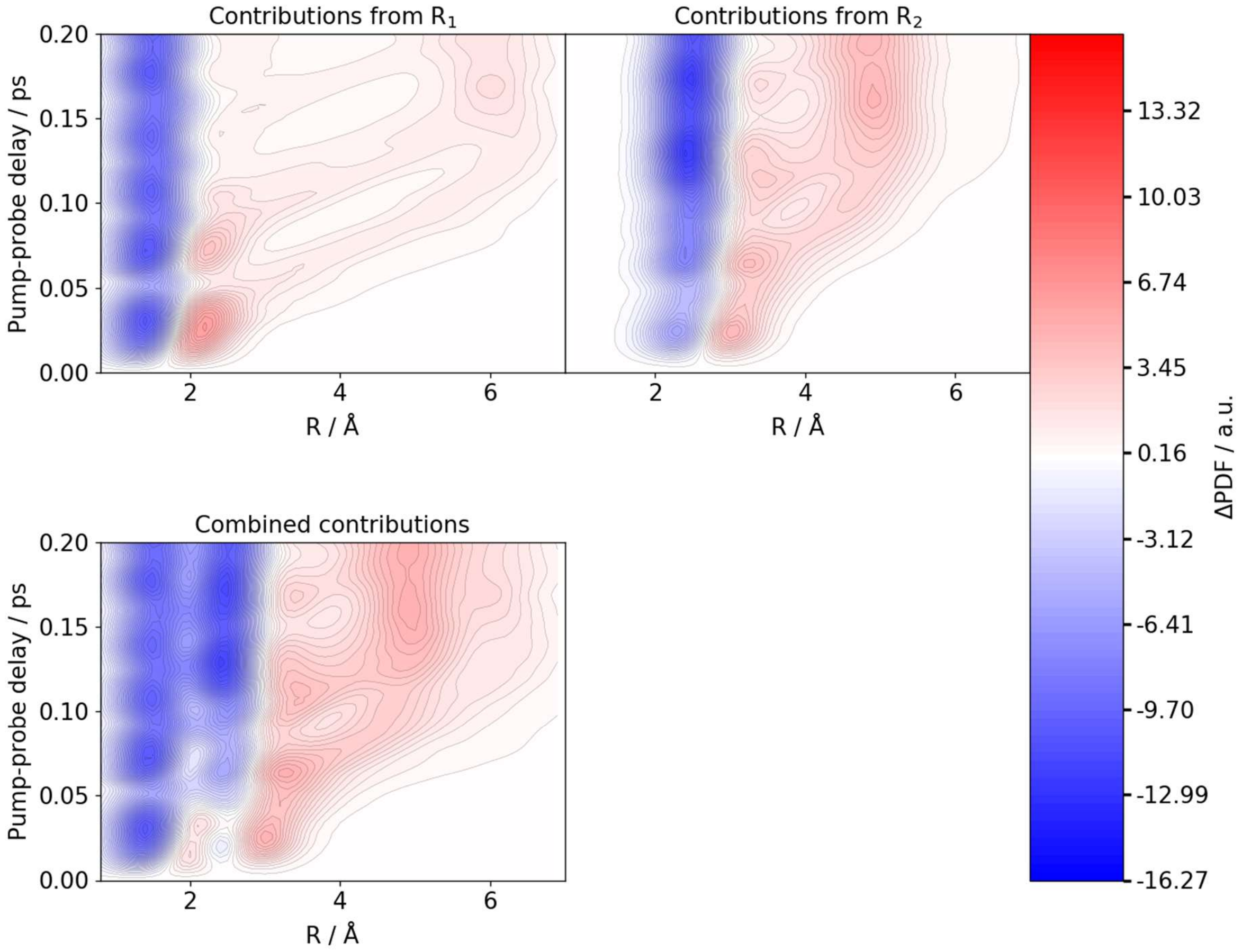

**Supplementary Figure 1: Visualization of the superposition of the contributions from simulated C-C distances $R_1$ and $R_2$ to the simulated ΔPDFs leading to a delay of the onset of peak β.** The contributions are computed from all 30 initial conditions of our *ab-initio* multiple spawning simulations. In contrast to Figure 3, the simulation results are not convolved with a temporal response function. The initial negative contribution from $R_2$ due to bleaching of the $R_2$ value in ground state cyclohexadiene is efficiently compensated by the positive contribution from $R_1$.

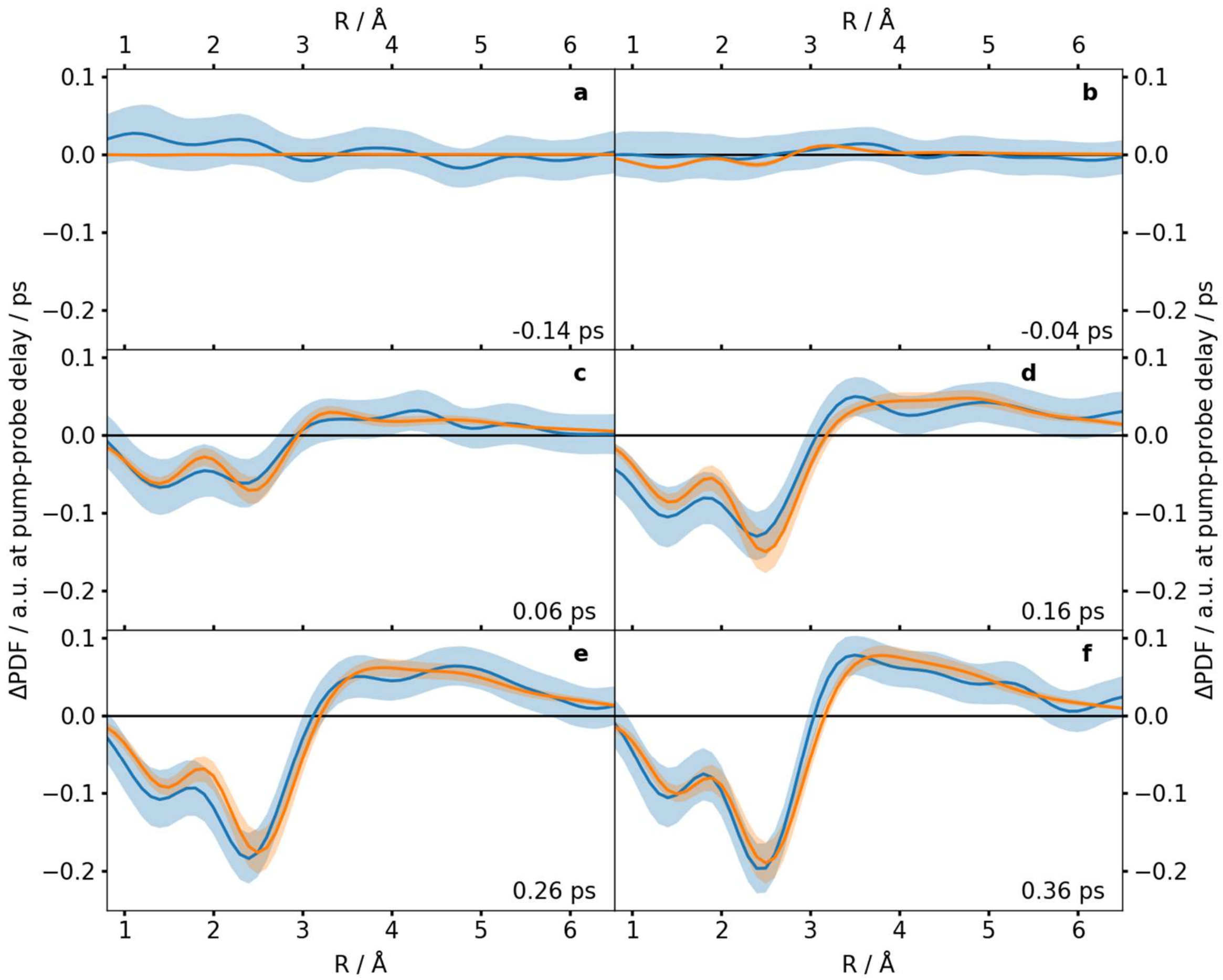

**Supplementary Figure 2: Direct comparison of experimental (blue) and simulated difference pair distribution functions (ΔPDF).** a-f) ΔPDFs at different pump-probe delay times (noted in the lower right corner). Error bars represent a 68 % confidence interval obtained from bootstrap analysis.[34] For the simulations, these error bars reflect convergence with respect to initial condition sampling.

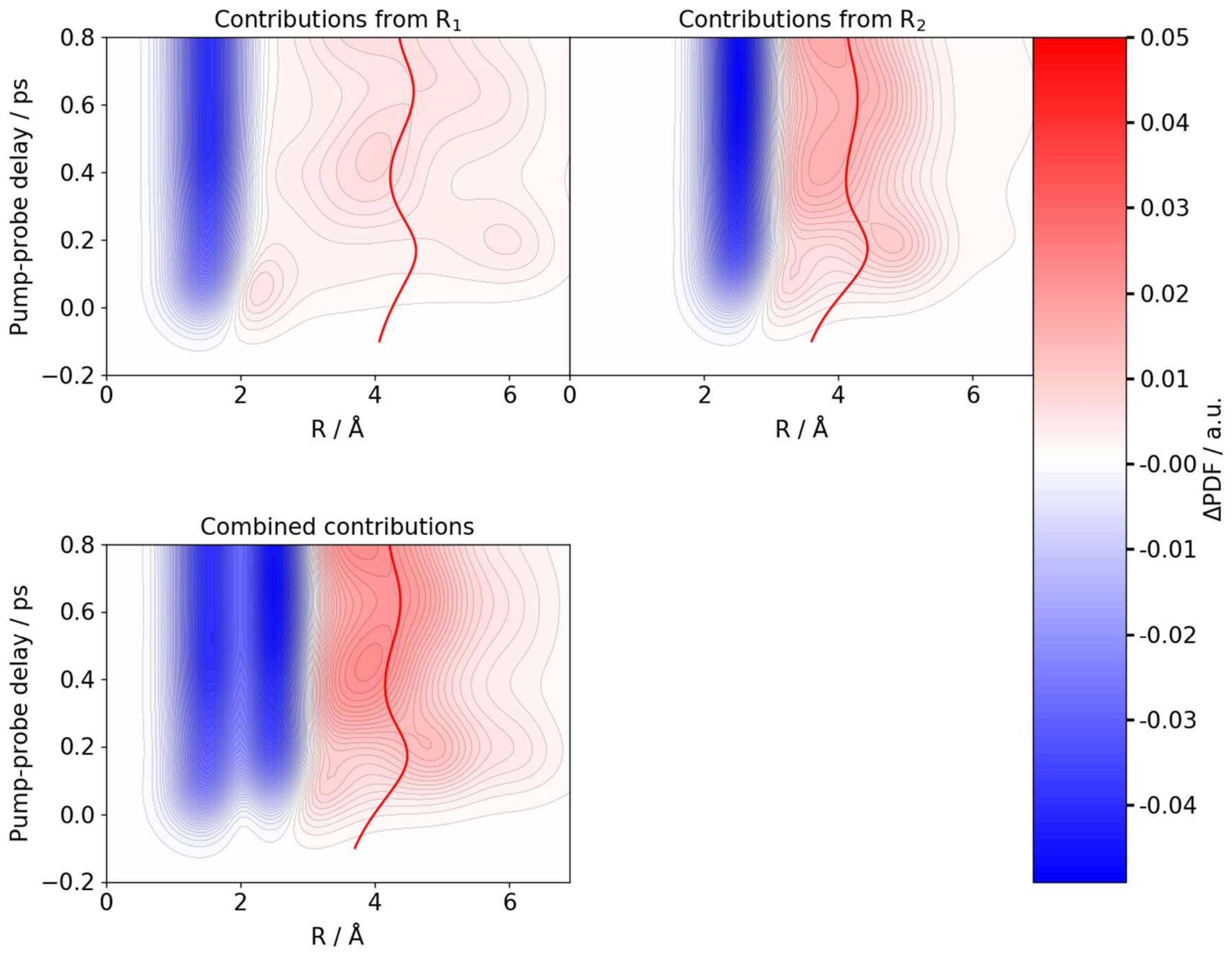

**Supplementary Figure 3: Visualization of the superposition of the contributions from simulated C-C distances $R_1$ and $R_2$ leading to the oscillatory feature in the peak γ region.** The contributions are computed from all 30 initial conditions of our *ab-initio* multiple spawning simulations and convolved with a temporal response function. The red curves represent the center-of-mass position in the peak γ region analog to Figure 3.

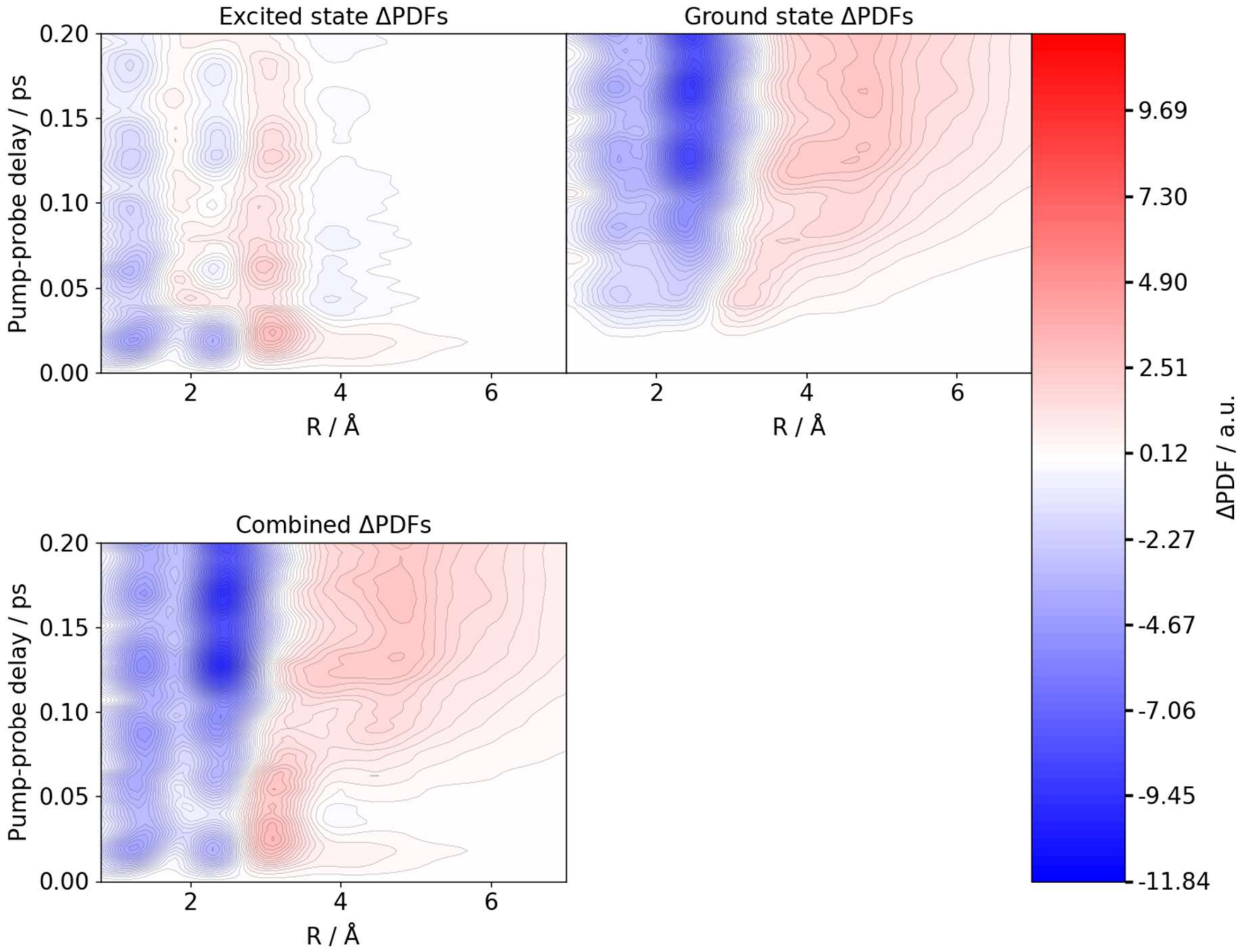

**Supplementary Figure 4: Excited/ground state resolved ΔPDFs.** AIMS-calculated excited state, ground state, and combined contributions to difference pair distribution functions (ΔPDFs) for the first 200fs after photoexcitation to $S_1$.

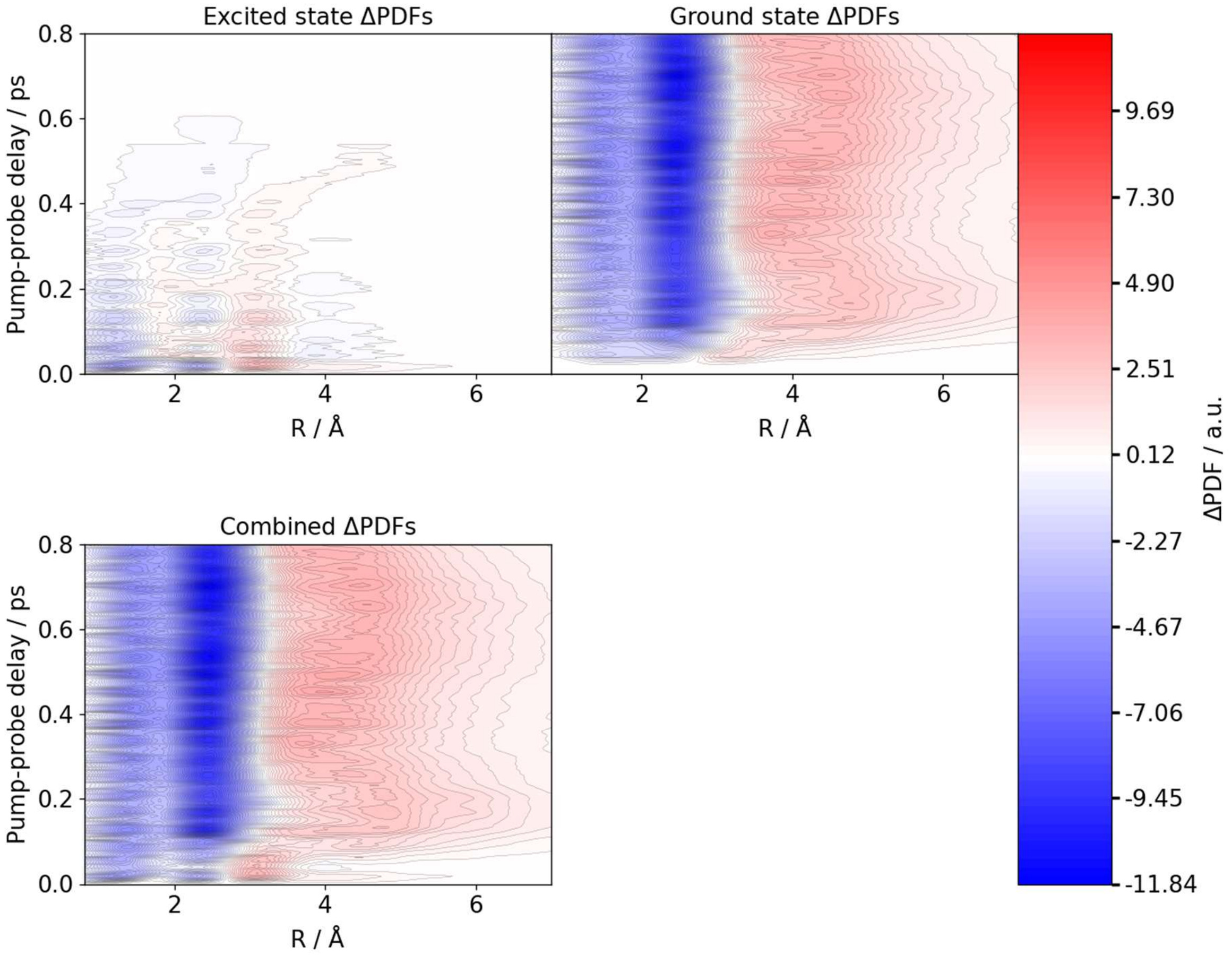

**Supplementary Figure 5: Excited/ground state resolved ΔPDFs.** AIMS-calculated excited state, ground state, and combined contributions to difference pair distribution functions (ΔPDFs) for the first 1ps after photoexcitation to $S_1$. As in Supplementary Figure 4, but for a longer time scale to show the decay of excited state contributions as the population reverts to the ground state.

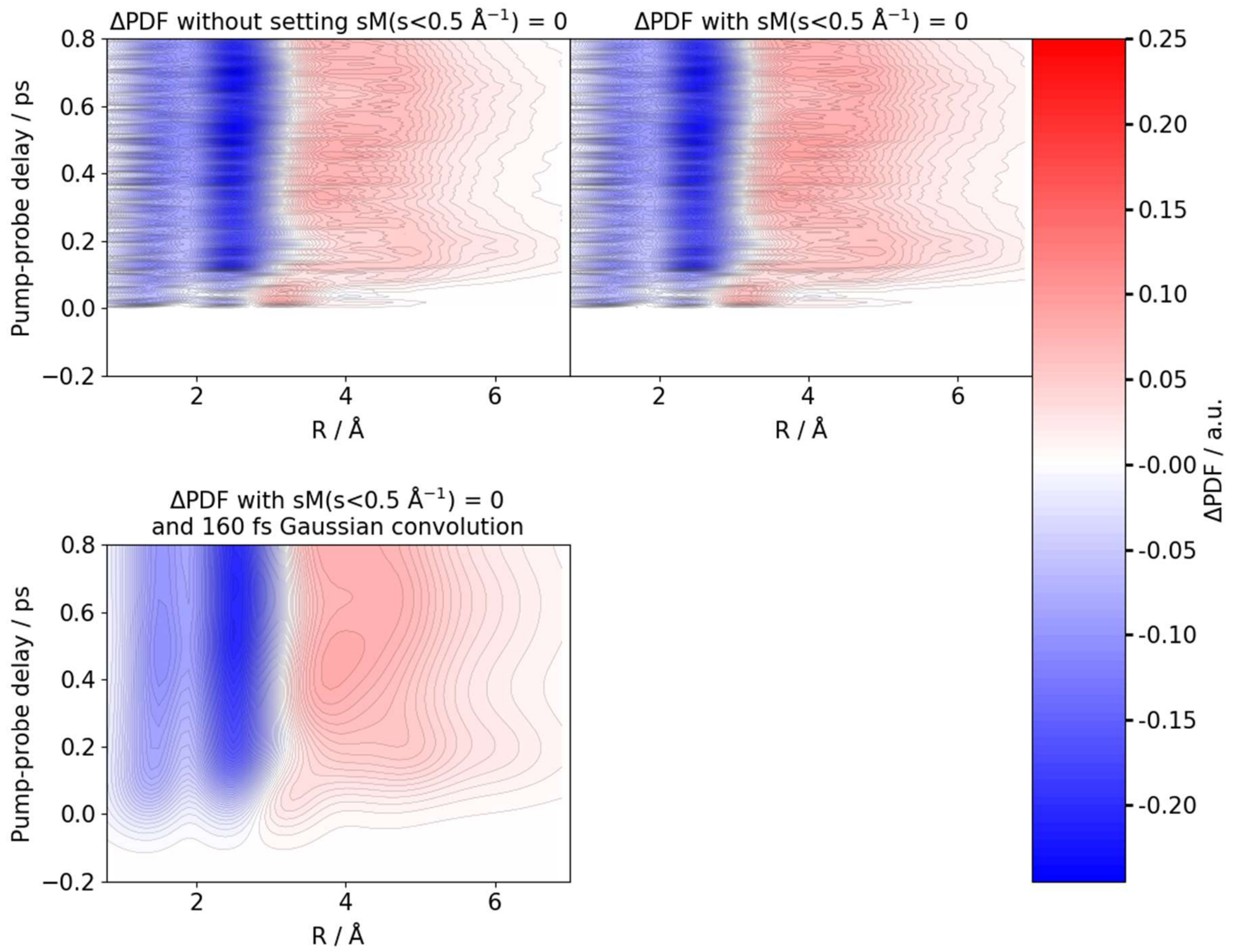

**Supplementary Figure 6: Theoretical ΔPDFs AIMS from simulations:** The theoretical ΔPDFs computed from the AIMS trajectories with sM(s < 0.5 Å $^{-1}$) not set to zero (left) and set to zero (right). The same ΔPDF convolved with a 160fs Gaussian (bottom).

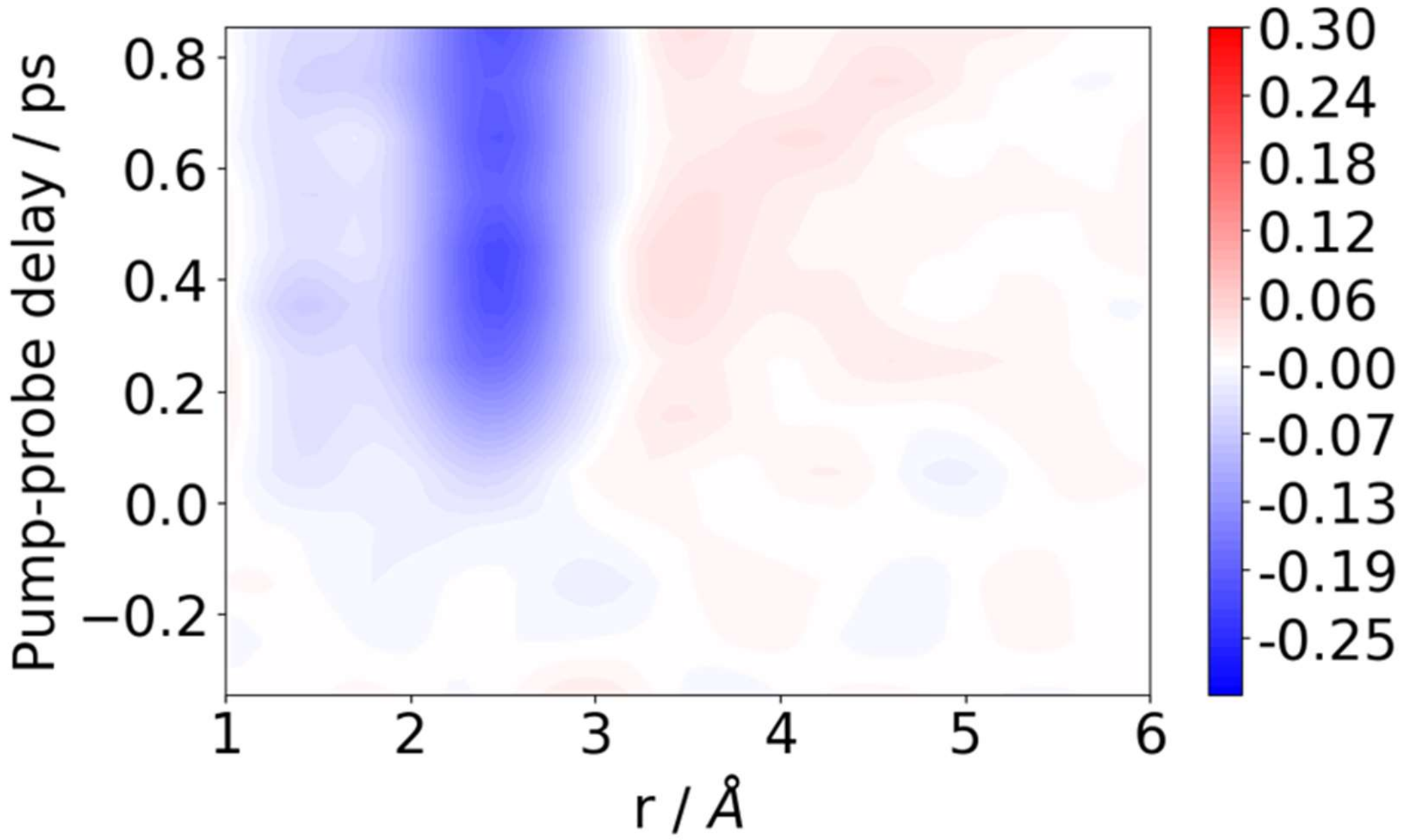

**Supplementary Figure 7: Experimental ΔPDF:** False color plot of experimental difference pair distribution functions resulting from setting molecular diffraction for s < 1.3 $Å^{-1}$ to zero

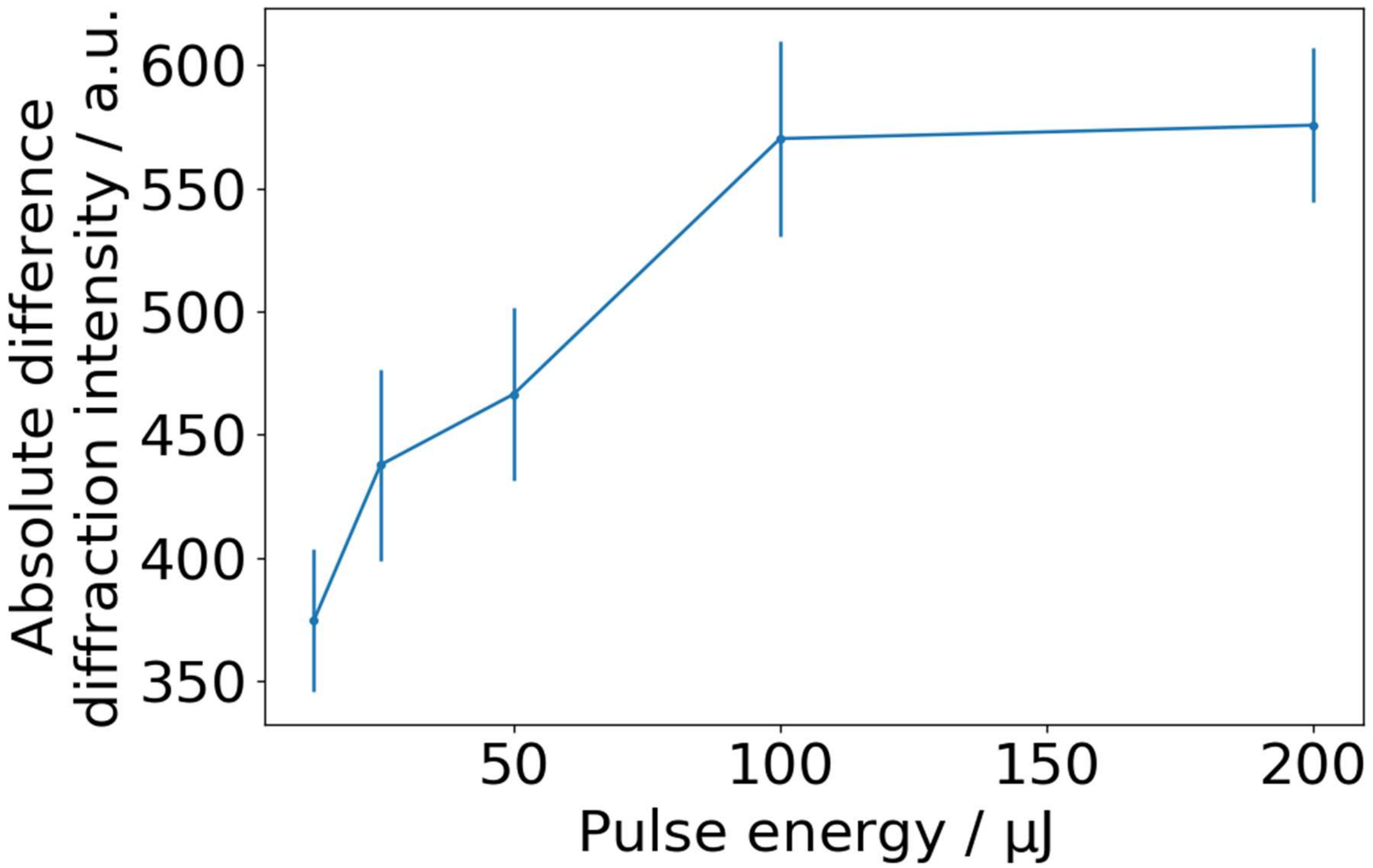

**Supplementary Figure 8: Results of a pump pulse intensity scan.** Error bars represent a 68 % confidence interval obtained from bootstrap analysis.[34]

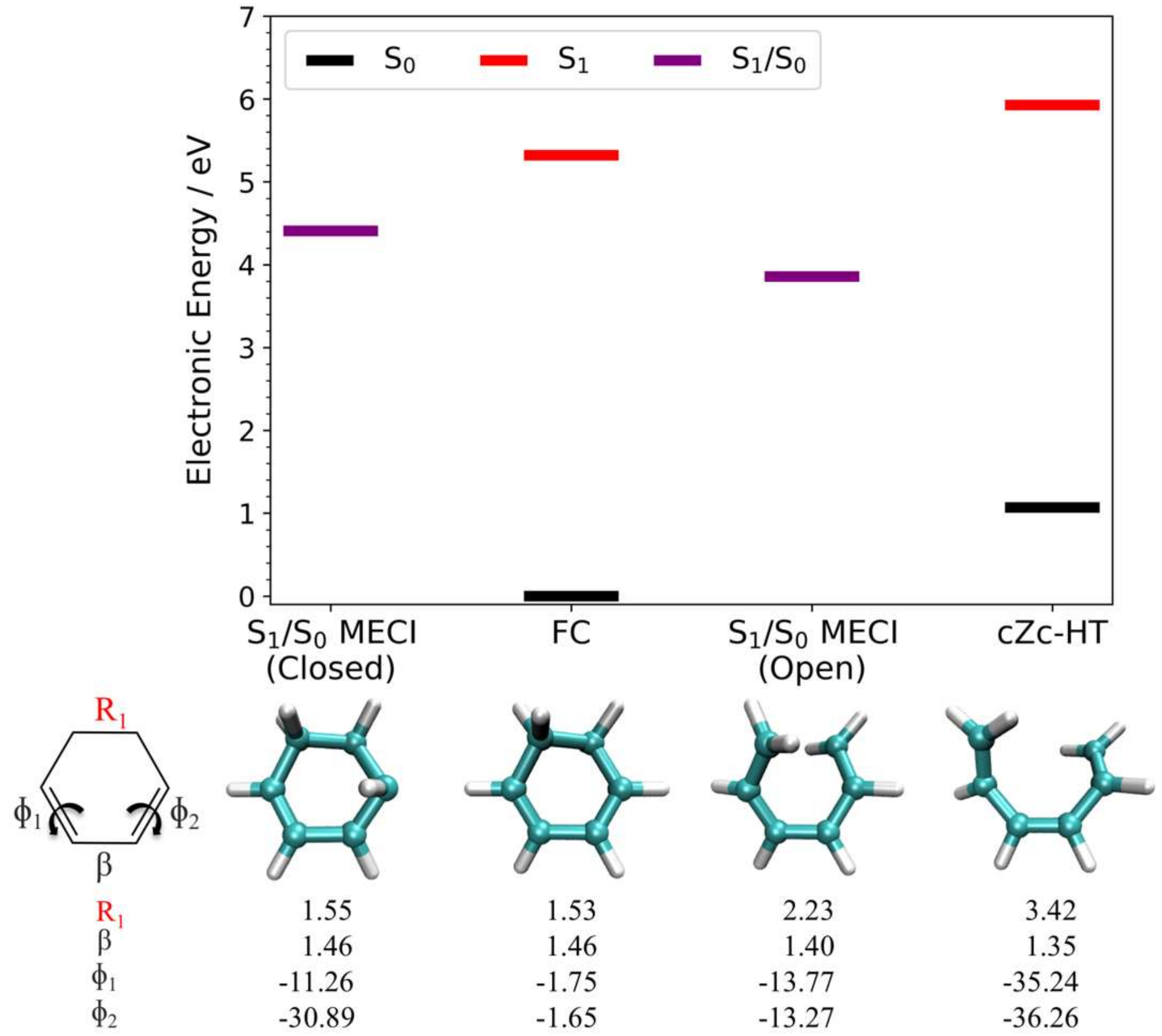

**Supplementary Figure 9: Critical points along the ring-opening pathway of CHD.** All energies are computed at the α-SA-2-CASSCF(6,4)/6-31G* level of theory with reference to the ground state energy of the Frank Condon (FC) point. Bond distances (Å) and dihedral angles (degrees) are shown under each structure.

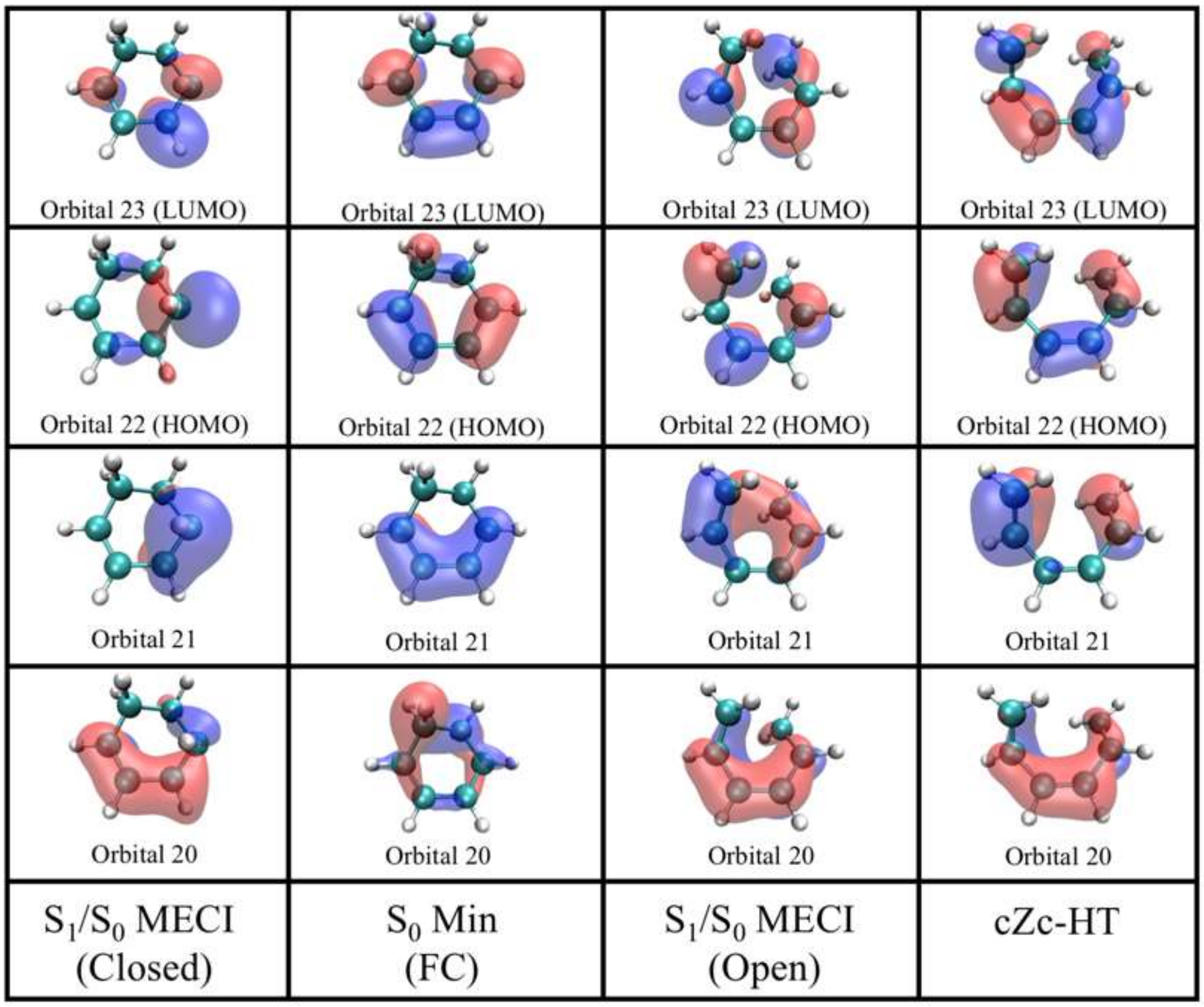

**Supplementary Figure 10: Molecular orbitals for critical points along the ring-opening pathway of CHD.** The active space orbitals for the critical points in Supplementary Figure 1. Blue and red correspond to positive and negative isovalues, respectively.

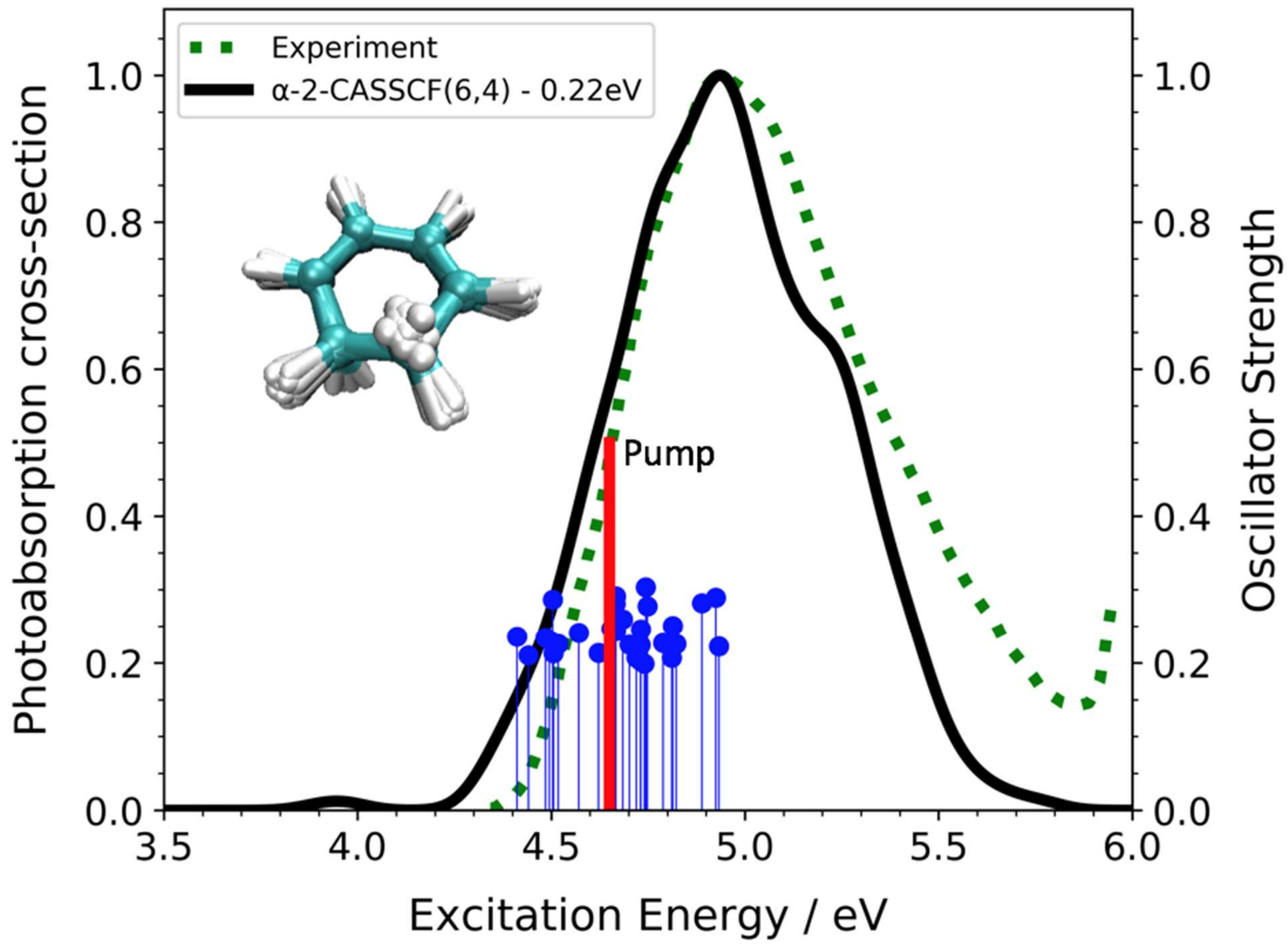

**Supplementary Figure 11: UV electronic absorption spectrum:** The UV electronic absorption spectrum was generated from 500 initial conditions sampled from a ground state harmonic Wigner distribution. The oscillator strengths of the 30 initial conditions used for the AIMS simulations (blue) are chosen ± 0.30 eV around the central wavelength of the pump pulse used in the UED experiment.

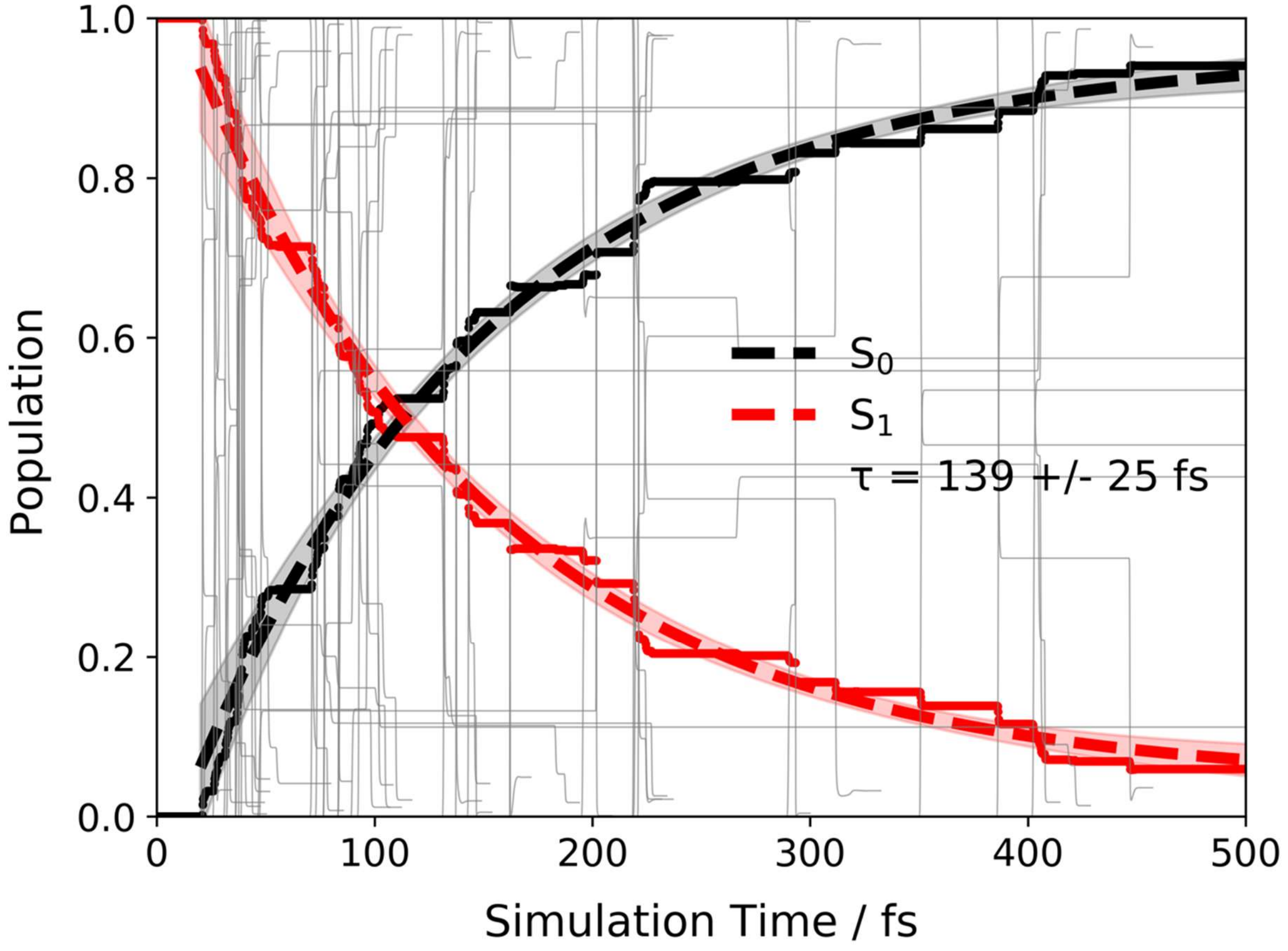

**Supplementary Figure 12: Population Dynamics for CHD:** The total population ($S_1$-red and $S_0$-black) for all 30 initial conditions in the first 500fs of the AIMS simulations. Both $S_0$ and $S_1$ were fit to monoexponential curves and their associated uncertainty represents a 99% confidence interval obtained from bootstrap analysis. The grey lines represent the population for individual initial conditions.

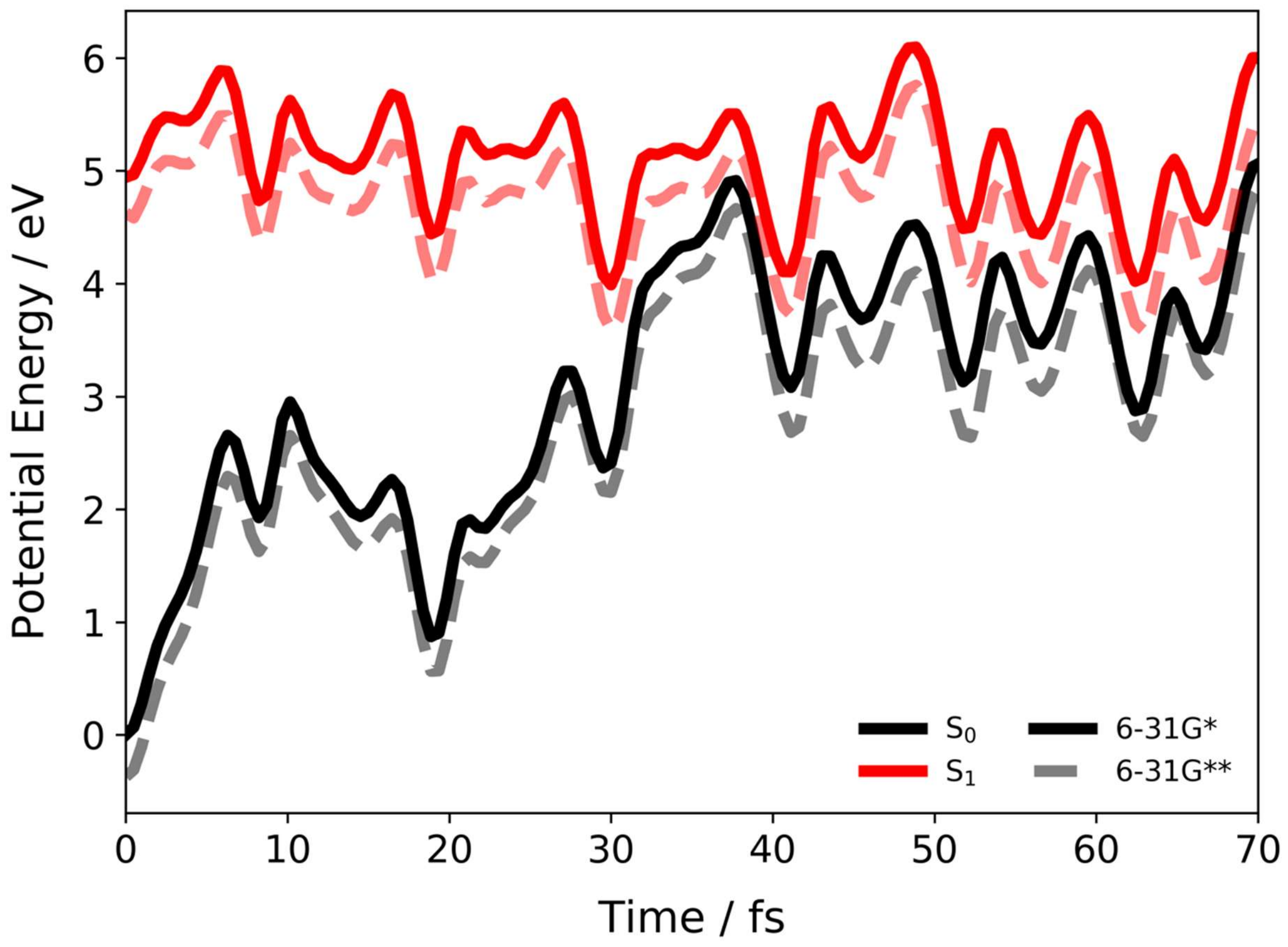

**Supplementary Figure 13: Impact of basis set size on excited state dynamics:** Potential energies of the two lowest singlet states ($S_1$ and $S_2$) of a trajectory basis function propagating on the $S_1$ state calculated using two different basis sets (6-31G* and 6-31G**). Only a small systematic shift is observed between the two basis sets.

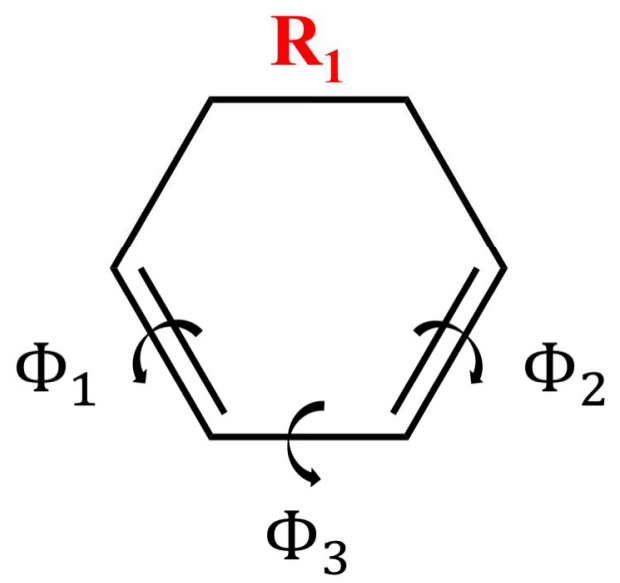

| | | | | |
|---|---|---|---|---|
| **CHD** | $R_1 \leq 1.8$ Å | $\|\Phi_1\| \leq 45^\circ$ | $\|\Phi_2\| \leq 45^\circ$ | $\|\Phi_3\| \leq 45^\circ$ |
| **cZc - HT** | $R_1 > 1.8$ Å | $\|\Phi_1\| \leq 45^\circ$ | $\|\Phi_2\| \leq 45^\circ$ | $\|\Phi_3\| \leq 45^\circ$ |
| **cZt - HT** | $R_1 > 1.8$ Å | $\|\Phi_1\| \leq 45^\circ$<br>$\|\Phi_1\| \geq 135^\circ$ | $\|\Phi_2\| \geq 135^\circ$<br>$\|\Phi_2\| \leq 45^\circ$ | $\|\Phi_3\| \leq 45^\circ$ |
| **tZt - HT** | $R_1 > 1.8$ Å | $\|\Phi_1\| \geq 135^\circ$ | $\|\Phi_2\| \geq 135^\circ$ | $\|\Phi_3\| \leq 45^\circ$ |

**Supplementary Figure 14: Classification of CHD and HT isomers on $S_0$ :** Each row corresponds to the four classification criteria used to bin geometries from the ground state TBF trajectories.

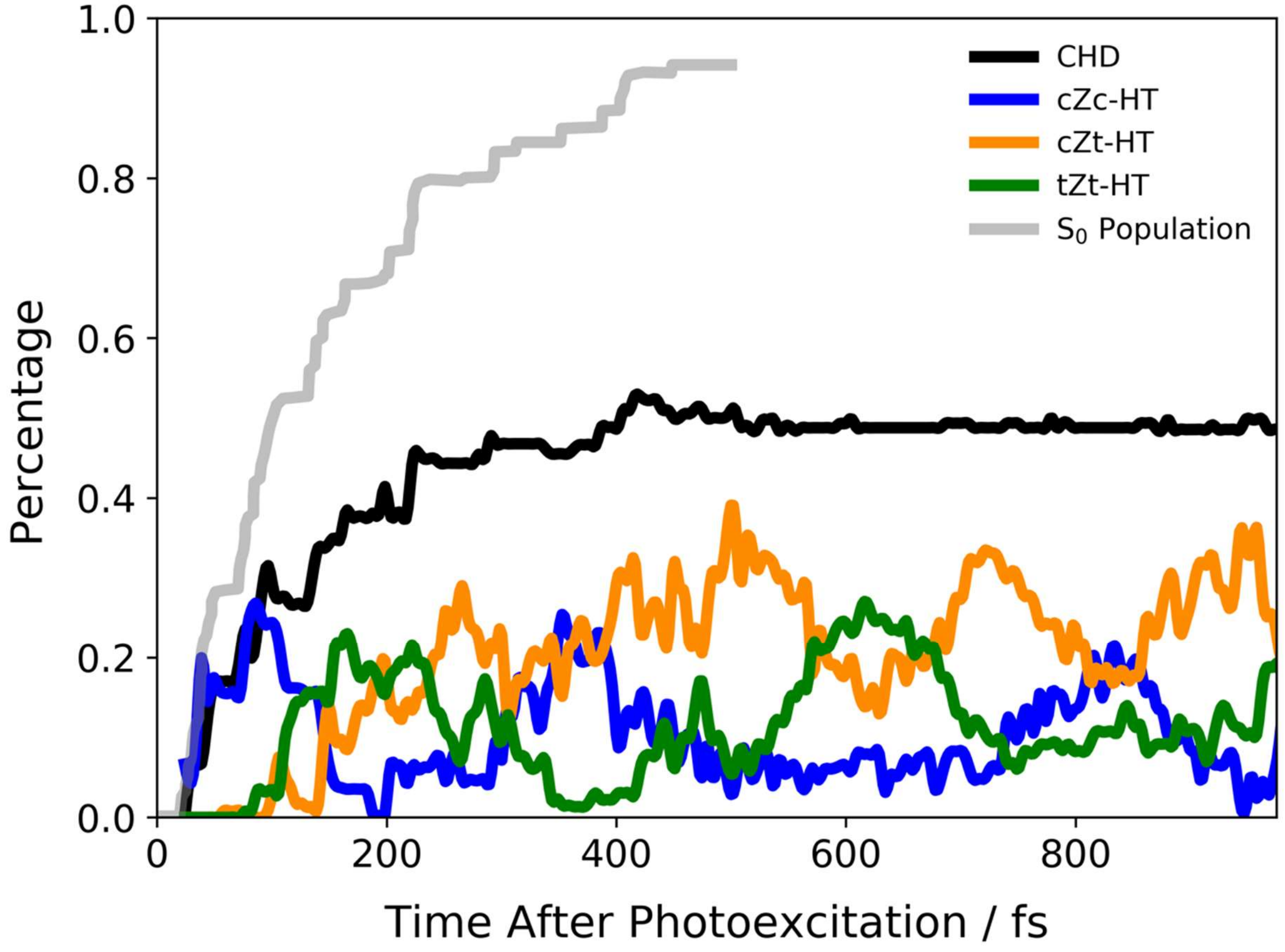

**Supplementary Figure 15: Population of CHD and HT isomers on $S_0$ after photoexcitation:** The percentage of the ground-state population binned into HT isomers from optimized ground-state geometries via torsional angles $\Phi_1$, $\Phi_2$, $\Phi_3$, and $R_1$ (Supplementary Figure 14). Time zero corresponds to the initial photoexcitation of the wavepacket to $S_1$.

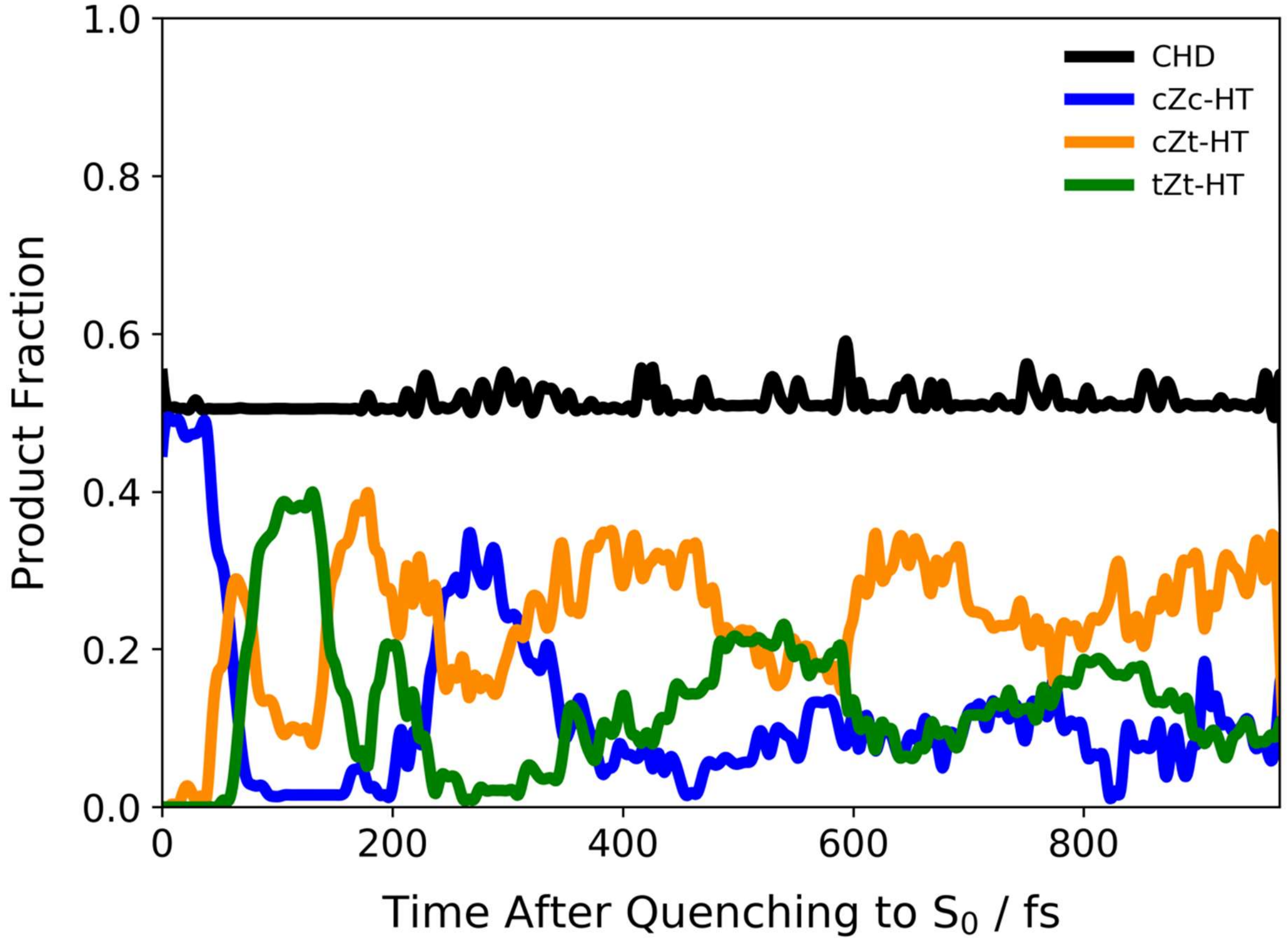

**Supplementary Figure 16: Population of HT isomers on the $S_0$ after quenching to $S_0$:** The percentage of the ground-state population binned into HT isomers from optimized ground-state geometries via torsional angles $\Phi_1$, $\Phi_2$, $\Phi_3$, and $R_1$ (Supplementary Figure 14). Time zero corresponds to the spawn time of each TBF to $S_0$.

## Supplementary Tables

| | | |
|---|---|---|
| | **$S_0$ Minimum (FC)** | |
| $S_0$ energy / H | -231.82028796441909 | |
| $S_0$ CI eigenvector | -0.98941257270382 | X20 X21 X22 |
| | 0.11495279788797 | X20 X21 X23 |
| | 0.07494922766996 | X20 X22 X23 |
| | -0.03181677028755 | A20 X21 B22 X23 |
| | -0.03181677028755 | B20 X21 A22 X23 |
| $S_1$ energy / H | -231.62477535476674 | |
| $S_1$ CI eigenvector | -0.70045437789207 | X20 X21 A22 B23 |
| | -0.70045437789207 | X20 X21 B22 A23 |
| | 0.08402691521916 | X20 A21 B22 X23 |
| | 0.08402691521916 | X20 B21 A22 X23 |
| | -0.04294646022130 | A20 X21 X22 B23 |
| | -0.04294646022130 | B20 X21 X22 A23 |
| | -0.02141801381956 | A20 B21 X22 X23 |
| | -0.02141801381956 | B20 A21 X22 X23 |
| | **$S_1$/$S_0$ MECI-Closed** | |
| $S_0$ energy / H | -231.65827064523486 | |
| $S_0$ CI eigenvector | 0.93023211064854 | X20 X21 X22 |
| | -0.20949005638975 | X20 X21 A22 B23 |
| | -0.20949005638975 | X20 X21 B22 A23 |
| | 0.11895203133623 | X20 A21 X22 B23 |
| | 0.11895203133623 | X20 B21 X22 A23 |
| | -0.10822967370534 | X20 X22 X23 |
| | -0.05423452480264 | X21 X22 X23 |
| | 0.02793085359266 | X20 A21 B22 X23 |
| | 0.02793085359266 | X20 B21 A22 X23 |
| | 0.02107419485034 | A20 X21 X22 B23 |
| | 0.02107419485034 | B20 X21 X22 A23 |
| $S_1$ energy / H | -231.65827060674511 | |
| $S_1$ CI eigenvector | 0.66483473217041 | X20 X21 A22 B23 |
| | 0.66483473217041 | X20 X21 B22 A23 |
| | 0.29324575590757 | X20 X21 X22 |
| | -0.10345257179471 | X20 A21 B22 X23 |
| | -0.10345257179471 | X20 B21 A22 X23 |
| | -0.04713906229724 | A20 X21 B22 X23 |
| | -0.04713906229724 | B20 X21 A22 X23 |
| | 0.03541813022859 | X20 A21 X22 B23 |
| | 0.03541813022859 | X20 B21 X22 A23 |
| | -0.03337145560651 | X20 X22 X23 |
| | -0.02214337265540 | X21 X22 X23 |

| S$_1$/S$_0$ MECI-Open | | |
|---|---|---|
| S$_0$ energy / H | -231.67847837132669 | |
| S$_0$ CI eigenvector | -0.65239330226724 | X20 X21 A22 B23 |
| | -0.65239330226724 | X20 X21 B22 A23 |
| | -0.36360943529701 | X20 X21 X22 |
| | 0.06888073954829 | X20 X22 X23 |
| | 0.06243313379458 | X20 A21 B22 X23 |
| | 0.06243313379458 | X20 B21 A22 X23 |
| | -0.04755117829995 | X20 X21 X23 |
| | 0.03207678166485 | X21 X22 X23 |
| S$_1$ energy / H | -231.67847834689633 | |
| S$_1$ CI eigenvector | -0.90836147658198 | X20 X21 X22 |
| | 0.26389087198903 | X20 X21 A22 B23 |
| | 0.26389087198903 | X20 X21 B22 A23 |
| | 0.12804542332827 | X20 X22 X23 |
| | -0.08728387865539 | X20 X21 X23 |
| | 0.08461510236361 | X21 X22 X23 |
| | -0.03821939355837 | X20 A21 X22 B23 |
| | -0.03821939355837 | X20 B21 X22 A23 |
| | -0.02451304924853 | X20 A21 B22 X23 |
| | -0.02451304924853 | X20 B21 A22 X23 |

| HT-cZc Minimum | | |
|---|---|---|
| S$_0$ energy / H | -231.78098258009587 | |
| S$_0$ CI eigenvector | -0.98872815589489 | X20 X21 X22 |
| | 0.09865082052647 | X20 X21 X23 |
| | 0.07016497560625 | X21 X22 X23 |
| | 0.05116431220264 | A20 X21 B22 X23 |
| | 0.05116431220264 | B20 X21 A22 X23 |
| | 0.04865878772002 | X20 X22 X23 |
| S$_1$ energy / H | -231.60245424156005 | |
| S$_1$ CI eigenvector | -0.69213305311113 | X20 X21 A22 B23 |
| | -0.69213305311113 | X20 X21 B22 A23 |
| | 0.10331194051980 | X20 A21 B22 X23 |
| | 0.10331194051980 | X20 B21 A22 X23 |
| | 0.10118686070543 | A20 X21 X22 B23 |
| | 0.10118686070543 | B20 X21 X22 A23 |

**Supplementary Table 1: Energies and CI eigenvectors for S$_0$ and S$_1$ at for critical points on the ring opening pathway of 1,3-cyclohexadiene.** For the CI eigenvectors, XYY indicates that the YYth molecular orbital is doubly occupied, and AYY/BYY indicate that the YYth molecular orbital is singly occupied with alpha or beta spin, respectively.

## Supplementary Movies

### 1. Supplementary Movie 1:

**CHD formation through the open-ring CI.** This movie shows the elongation of $R_1$ as CHD propagates along the $S_1$ surface. After approximately 41fs, the nuclear wavepacket reaches the $S_1/S_0$ Open-CI, where $R_1$ increases to ~2.15Å and $R_2$ seems to change little. CHD relaxes back to $S_0$ via this open-ring CI, forming only CHD. (open-CI-0005-CHD.mp4)

### 2. Supplementary Movie 2:

**HT formation through the open-ring CI.** This movie shows the elongation of $R_1$ as CHD propagates along the $S_1$ surface. A substantial speed-up of the ring-opening can be observed upon returning to $S_0$. Furthermore, the rotation of the terminal double bonds leading to the coherent oscillations in the experimental **ΔPDF** can be observed. (open-CI-0100-HT.mp4)

### 3. Supplementary Movie 3:

**CHD formation through the closed-ring CI.** This movie shows the nonradiative relaxation pathway of CHD through the closed-ring CI. After approximately 180fs, CHD reaches the closed-ring CI, where CHD relaxes back to $S_0$, forming strictly CHD. Unlike the trajectories that pass through the open-CI, $R_1$ and $R_2$ stay relatively constant throughout the entire simulation. (closed-CI-0004-CHD.mp4)

## References

1 Carreira, L. A., Carter, R. O. & Durig, J. R. Raman spectra of gases. VII. Barriers to planarity in 1,4- and 1,3-cyclohexadiene. *J. Chem. Phys.* **59**, 812-816, (1973).

2 Salvat, F., Jablonski, A. & Powell, C. J. ELSEPA—Dirac partial-wave calculation of elastic scattering of electrons and positrons by atoms, positive ions and molecules. *Comp. Phys. Comm.* **165**, 157-190, (2005).

3 Firefly version 8.1.1 http://classic.chem.msu.su/gran/firefly/index.html v. 8.1.1.

4 Schmidt, M. W., Baldridge, K. K., Boatz, J. A., Elbert, S. T., Gordon, M. S., Jensen, J. H., Koseki, S., Matsunaga, N., Nguyen, K. A., Su, S., Windus, T. L., Dupuis, M. & Montgomery, J. A. General atomic and molecular electronic structure system. *J. Comput. Chem.* **14**, 1347-1363, (1993).

5 Ben-Nun, M. & Martínez, T. J. Nonadiabatic molecular dynamics: Validation of the multiple spawning method for a multidimensional problem. *J. Chem. Phys.* **108**, 7244-7257, (1998).

6 Ben-Nun, M., Quenneville, J. & Martínez, T. J. Ab Initio Multiple Spawning: Photochemistry from First Principles Quantum Molecular Dynamics. *J. Phys. Chem. A* **104**, 5161-5175, (2000).

7 Ben-Nun, M. & Martinez, T. J. Ab Initio Quantum Molecular Dynamics. *Adv. Chem. Phys.* **121**, 439-512, (2002).

8 Yang, S. & Martínez, T. J. in *Conical Intersections* 347-374 (WORLD SCIENTIFIC, 2012).

9 Hohenstein, E. G., Luehr, N., Ufimtsev, I. S. & Martinez, T. J. An atomic orbital-based formulation of the complete active space self-consistent field method on graphical processing units. *J Chem Phys* **142**, 224103, (2015).

10 Snyder, J. W., Hohenstein, E. G., Luehr, N. & Martínez, T. J. An atomic orbital-based formulation of analytical gradients and nonadiabatic coupling vector elements for the state-averaged complete active space self-consistent field method on graphical processing units. *J. Chem. Phys.* **143**, 154107, (2015).

11 Snyder, J. W., Fales, B. S., Hohenstein, E. G., Levine, B. G. & Martinez, T. J. A direct-compatible formulation of the coupled perturbed complete active space self-consistent field equations on graphical processing units. *J Chem Phys* **146**, 174113, (2017).

12 Snyder, J. W., Curchod, B. F. E. & Martínez, T. J. GPU-Accelerated State-Averaged Complete Active Space Self-Consistent Field Interfaced with Ab Initio Multiple Spawning Unravels the Photodynamics of Provitamin D3. *J. Phys. Chem. Lett.* **7**, 2444-2449, (2016).

13 Snyder, J. W., Parrish, R. M. & Martínez, T. J. α-CASSCF: An Efficient, Empirical Correction for SA-CASSCF To Closely Approximate MS-CASPT2 Potential Energy Surfaces. *J. Phys. Chem. Lett.* **8**, 2432-2437, (2017).

14 Arruda, B. C. & Sension, R. J. Ultrafast polyene dynamics: the ring opening of 1,3-cyclohexadiene derivatives. *Phys. Chem. Chem. Phys.* **16**, 4439-4455, (2014).

15 Deb, S. & Weber, P. M. The Ultrafast Pathway of Photon-Induced Electrocyclic Ring-Opening Reactions: The Case of 1,3-Cyclohexadiene. *Ann. Rev. Phys. Chem.* **62**, 19-39, (2011).

16 Tao, H. L. *First Principles Molecular Dynamics and Control of Photochemical Reactions* PhD thesis, Stanford University, (2011).

17 Kosma, K., Trushin, S. A., Fuss, W. & Schmid, W. E. Cyclohexadiene ring opening observed with 13 fs resolution: coherent oscillations confirm the reaction path. *Phys. Chem. Chem. Phys.* **11**, 172-181, (2009).

18 Fuß, W., Schmid, W. E. & Trushin, S. A. Time-resolved dissociative intense-laser field ionization for probing dynamics: Femtosecond photochemical ring opening of 1,3-cyclohexadiene. *J. Chem. Phys.* **112**, 8347-8362, (2000).

19 Ufimtsev, I. S. & Martínez, T. J. Quantum Chemistry on Graphical Processing Units. 1. Strategies for Two-Electron Integral Evaluation. *J. Chem. Theo. Comp.* **4**, 222-231, (2008).

20 Ufimtsev, I. S. & Martinez, T. J. Quantum Chemistry on Graphical Processing Units. 2. Direct Self-Consistent-Field Implementation. *J. Chem. Theo. Comp.* **5**, 1004-1015, (2009).

21 Ufimtsev, I. S. & Martinez, T. J. Quantum Chemistry on Graphical Processing Units. 3. Analytical Energy Gradients, Geometry Optimization, and First Principles Molecular Dynamics. *J. Chem. Theo. Comp.* **5**, 2619-2628, (2009).

22 Kästner, J., Carr, J. M., Keal, T. W., Thiel, W., Wander, A. & Sherwood, P. DL-FIND: An Open-Source Geometry Optimizer for Atomistic Simulations. *J. Phys. Chem. A* **113**, 11856-11865, (2009).

23 Curchod, B. F. E. & Martínez, T. J. Ab Initio Nonadiabatic Quantum Molecular Dynamics. *Chem. Rev.* **118**, 3305-3336, (2018).

24 Born, M. & Huang, K. *Dynamical Theory of Crystal Lattices*. (Clarendon Press, 1998).

25 Curchod, B. F. E., Sisto, A. & Martínez, T. J. Ab Initio Multiple Spawning Photochemical Dynamics of DMABN Using GPUs. *J. Phys. Chem. A* **121**, 265-276, (2017).

26 Kuthirummal, N., Rudakov, F. M., Evans, C. L. & Weber, P. M. Spectroscopy and femtosecond dynamics of the ring opening reaction of 1,3-cyclohexadiene. *J. Chem. Phys.* **125**, 133307, (2006).

27 Attar, A. R., Bhattacherjee, A., Pemmaraju, C. D., Schnorr, K., Closser, K. D., Prendergast, D. & Leone, S. R. Femtosecond x-ray spectroscopy of an electrocyclic ring-opening reaction. *Science* **356**, 54-59, (2017).

28 Tapavicza, E., Meyer, A. M. & Furche, F. Unravelling the details of vitamin D photosynthesis by non-adiabatic molecular dynamics simulations. *Phys. Chem. Chem. Phys.* **13**, 20986-20998, (2011).

29 Levine, B. G., Ko, C., Quenneville, J. & MartÍnez, T. J. Conical intersections and double excitations in time-dependent density functional theory. *Mol. Phys.* **104**, 1039-1051, (2006).

30 Schäfer, L. Electron Diffraction as a Tool of Structural Chemistry. *Appl. Spectr.* **30**, 123-149, (1976).

31 Thompson, A. L., Punwong, C. & Martínez, T. J. Optimization of width parameters for quantum dynamics with frozen Gaussian basis sets. *Chem. Phys.* **370**, 70-77, (2010).

32 Makhov, D. V., Glover, W. J., Martinez, T. J. & Shalashilin, D. V. Ab initio multiple cloning algorithm for quantum nonadiabatic molecular dynamics. *J. Chem. Phys.* **141**, 054110, (2014).

33 Kim, J., Tao, H., Martinez, T. J. & Bucksbaum, P. Ab initio multiple spawning on laser-dressed states: a study of 1,3-cyclohexadiene photoisomerization via light-induced conical intersections. *J. Phys. B* **48**, 164003, (2015).

34 Shao, J. & Tu, D. *The Jackknife and Bootstrap*. (Springer-Verlag, 1995).